\documentclass{jpp}
\usepackage{graphicx}
\usepackage{float}

 \usepackage{amssymb}
 \usepackage{bigints}
 \usepackage{amsmath}
 \usepackage{empheq}
 \usepackage{framed}
 \usepackage{color}
 \usepackage{enumerate}
 \usepackage{ulem}
 \usepackage{bm}
 \usepackage{appendix}
 \usepackage{multicol}
 \numberwithin{equation}{section}
 \usepackage{upgreek}
 \usepackage{textgreek}
 \usepackage{graphicx}
\usepackage{stmaryrd} 
\usepackage{xfrac}    

\newcommand{\comment}[1]{}

\newcommand{\be}{\begin{equation}}
\newcommand{\ee}{\end{equation}}
\newcommand{\ba}{\[\begin{aligned}}
\newcommand{\ea}{\end{aligned}\]}
\newcommand{\bea}{\begin{eqnarray}}
\newcommand{\eea}{\end{eqnarray}}
\newcommand{\beann}{\begin{eqnarray*}}
\newcommand{\eeann}{\end{eqnarray*}}
\newcommand{\bs}{\begin{split}}
\newcommand{\es}{\end{split}}
\renewcommand{\Re}{\operatorname{Re}}

\newcommand*{\cA}{\mathcal{A}} 

\newcommand*{\cC}{\mathcal{C}}
\newcommand*{\cD}{\mathcal{D}}
\newcommand*{\cE}{\mathcal{E}}
\newcommand*{\cF}{\mathcal{F}}
\newcommand*{\cG}{\mathcal{G}}
\newcommand*{\cH}{\mathcal{H}}

\newcommand*{\cJ}{\mathcal{J}}
\newcommand*{\cK}{\mathcal{K}}
\newcommand*{\cL}{\mathcal{L}}

\newcommand*{\mfc}{\mathfrak{c}}

\newcommand*{\ep}{\epsilon}
\newcommand*{\vep}{\varepsilon}

\newcommand*{\B}{\bm{B}}

\newcommand*{\J}{\bm{J}}

\newcommand*{\w}{\bm{w}}

\newcommand*{\BDB}{\B\cdot \dl B}

\newcommand*{\uD}{\u\cdot \dl }

\renewcommand*{\u}{\bm{u}}
\renewcommand*{\v}{\bm{v}}

\newcommand*{\jpl}{j_{||}}

\newcommand*{\dz}{\partial_z}
\newcommand*{\dR}{\partial_R}
\newcommand*{\dr}{\partial_r}

\newcommand*{\dl}{\bm{\nabla}}
\newcommand*{\del}{\partial}
\newcommand*{\BD}{\bm{B}\cdot\bm{\nabla}}

\newcommand*{\dlts}{\Delta^*}

\newcommand*{\zh}{\bm{\hat{z}}}

\newcommand*{\Rh}{\bm{\hat{R}}}

\newcommand*{\lbr}{\left(}
\newcommand*{\rbr}{\right)}

\newcommand*\tnsr[1]{\mathsfbi{#1}}

\newcommand*\at[2]{\left.#1\right|_{#2}}

\newcommand*\reallywidetilde[1]{\ThisStyle{%
  \setbox0=\hbox{$\SavedStyle#1$}%
  \stackengine{-.1\LMpt}{$\SavedStyle#1$}{%
    \stretchto{\scaleto{\SavedStyle\mkern.2mu\sim}{.5467\wd0}}{.7\ht0}%
  }{O}{c}{F}{T}{S}%
}}

\def\test#1{$%
  \reallywidetilde{#1}\,
  \scriptstyle\reallywidetilde{#1}\,
  \scriptscriptstyle\reallywidetilde{#1}
$\par}
\makeatletter
\newsavebox{\@brx}
\newcommand{\llangle}[1][]{\savebox{\@brx}{\(\m@th{#1\langle}\)}%
  \mathopen{\copy\@brx\mkern2mu\kern-0.9\wd\@brx\usebox{\@brx}}}
\newcommand{\rrangle}[1][]{\savebox{\@brx}{\(\m@th{#1\rangle}\)}%
  \mathclose{\copy\@brx\mkern2mu\kern-0.9\wd\@brx\usebox{\@brx}}}
\makeatother

\shorttitle{Compact NASE for QAS} 

\title{Compact quasiaxisymmetric stellarators, a near axisymmetric theory }

\shortauthor{W. Sengupta et al.}
\author{Wrick Sengupta\aff{1}\corresp{\email{wsengupta@princeton.edu}}, Rogerio Jorge \aff{2},  Nikita Nikulsin\aff{3}, Stefan Buller\aff{1}, Richard Nies \aff{4}, Andrew Brown \aff{1}, Amitava Bhattacharjee\aff{1}}

 \affiliation{
 \aff{1} Department of Astrophysical Sciences, Princeton University, Princeton, NJ 08543, USA
 \aff{2} Department of Physics, University of Wisconsin-Madison, Madison, Wisconsin 53706, USA
 \aff{3} Stellarator Theory Department, Max Plank Institute for Plasma Physics, 17491 Greifswald, Germany
 \aff{4} Rudolf Peierls Centre for Theoretical Physics, Parks Road, Oxford, OX1 3PU, UK
 }
\begin{document}

\maketitle

\begin{abstract}
We develop a theory of ridges in compact stellarators with quasiaxisymmetry (QA). The equilibrium with finite plasma currents and pressure is modeled by ideal magnetohydrostatics (MHS). Field lines are collimated near sharp ridges, much like X-points, making ridges attractive to divertor designs without the requirement of a rational rotational transform at the divertor. However, unlike X-points, which must cover the entire torus an integer number of times, sharp ridges are typically localized in certain parts of the flux surfaces. Motivated by recent work (Henneberg and Plunk, Phys. Rev. Research 6, L022052) on compact hybrid devices, we develop a perturbative treatment of nearly axisymmetric quasisymmetric devices by expanding in the deviation from perfect axisymmetry. As a result, we can analytically describe the key features of compact QA devices, such as the tendency for ridges to be localized on the inboard side, where the Gaussian curvature is typically negative, and the field strength is maximum. We provide comprehensive numerical evidence in support of our analytical theory.
\end{abstract}

\section{Introduction \label{sec:intro}}
Sharp ridges are an ubiquitous feature of most optimized stellarators (\cite{strumberger1996sol,nuhrenberg2006critical,boozer_2015}). These are typically localized regions of large principal curvature that appear as creases on the flux surfaces. They collimate magnetic fields along themselves and therefore could play a potentially important role in the design of nonresonant divertors \citep{bader2017_resilient_div_hsx,garcia2023explorationCTH,garcia2024resilient}. There is also significant flux expansion near the ridges. The gradient of the flux surface is minimum along the ridge and can vanish in the limit of infinitely large principal curvature associated with the ridge. For compact devices with quasiaxisymmetry (QA), the ridges are typically seen on the inboard side \citep{plunk2018,plunk2020_near_axisymmetry_MHD,henneberg2024compact,Schuett_henneberg2024compactQA,nikulsin2025a}, where the field strength is maximum, and the Gaussian curvature is non-positive. We developed an analytic theory of ridges for vacuum quasisymmetric devices in \cite{PaperI} hereafter referred to as Paper I, where we provided mathematical justifications for these properties.


Quasiaxisymmetric devices typically have substantial bootstrap currents \citep{landremanBullerDrevlak2022optimization}. Hence, a generalization to plasma equilibrium beyond vacuum fields is essential to allow further exploration of the interconnections between geometry, QA, and field strength. The robustness of the field-line focusing effects near the ridges to currents and finite plasma pressure requires careful investigation. 

Although the association of ridges with the maximum of a principal curvature and the alignment with the field line has been duly noted \citep{bader2017_resilient_div_hsx,boozer_2015}, the connection of ridges with Gaussian curvature has not been thoroughly explored in the stellarator literature. Uncovering the role of Gaussian curvature in a quasisymmetric MHS with finite beta and gaining analytical insight into this class of devices is the primary focus of this work. For large aspect ratio devices, expansion in the inverse aspect ratio provides intuition for both finite-beta tokamaks and stellarators \citep{strauss1997reduced_near_vacuum,freidberg2014idealMHD,sengupta2024QSHBS,nikulsin_sengupta2024GS,brown2025Palumbo}. For studying compact devices analytically, we restrict ourselves to nearly axisymmetric quasiaxisymmetric stellarators (QAS) in the following. The asymptotic expansion parameter is the deviation from axisymmetry, similar to \citet{plunk2018,plunk2020_near_axisymmetry_MHD}. The crucial difference in our approach from \citet{plunk2018} is that we enforce volumetric QA, whereas \citet{plunk2018} enforced QA on the outer boundary. 

In this work, we present a generalized Grad-Shafranov approach \citep{burby2020} to understanding the equilibrium of a plasma obeying magnetohydrostatic (MHS) force balance in a compact near-axisymmetric QAS \citep{plunk2020_near_axisymmetry_MHD} . We discuss ideal MHS and QS constraints in Section \ref{sec:QS_iMHS_basics}, and present the governing equations in Section \ref{sec:GS_picture}. We expand near axisymmetry in Section \ref{sec:NASE} with configurations like those of \citet{henneberg2024compact,Schuett_henneberg2024compactQA} in mind using the regular cylindrical $(R,\phi,z)$ coordinates. Consequently, the description of MHS force balance and the interplay between geometry and QA becomes more transparent, which we discuss in Section \ref{sec:summary}. We formulate the conditions under which ridges can be obtained in Section \ref{sec:Ridges_LargeN} and Section \ref{sec:bal_ridges} and obtain approximate analytical ridge-like solutions that explore different boundary conditions. Our key results show that non-resonant ridges in QAS prefer the inboard side, where the Gaussian curvature of the background axisymmetric flux surface is either zero or negative. The nonsymmetric perturbation vanishes on the outboard side. Finally, we present ample numerical verifications of the key analytical insights in Section \ref{sec:Numerical} and summarize our results in Section \ref{sec:conclusions}.

\section{Ideal MHS with the quasisymmetry constraint }\label{sec:QS_iMHS_basics}

We will assume ideal MHS equilibrium \citep{freidberg2014idealMHD,helander2014} described by
\begin{subequations}
\begin{align}
    \dl \cdot \B=0 \label{divB},\\
     \J\times \B = \dl p  \label{JxB},\\ 
     \J=\dl \times \B,  \label{JiscurlB}
\end{align}
    \label{eq:ideal_MHS}
\end{subequations}
as a model for the plasma equilibrium. Here, $\B,\J$, and $p$ denote the magnetic field vector, the current density, and the plasma pressure, respectively. We have used units such that $\mu_0=1$. Since we
will be looking at near-axisymmetric systems, it is convenient to assume that the pressure is constant on a set of nested flux surfaces labeled by a smooth \textit{poloidal} flux function $\psi$, such that
\begin{align}
    p=p(\psi), \quad \BD\psi=0, \quad \J\cdot \dl \psi=0.
    \label{eq:p_of_psi}
\end{align}
and that the gradient $|\dl \psi|$ is nonzero everywhere except for the magnetic axis.  

\subsection{Quasisymmetry}
We now further impose the constraint of quasisymmetry (QS). The so-called two-term form of QS \citep{helander2014,rodriguez2020a}, given by the following equation on the field strength $B$,
\begin{align}
    \B \times \dl \psi \cdot \dl B =F(\psi) \BDB
    \label{eq:2term_QS},
\end{align}
is one of the classic ways of encapsulating QS. For quasi-axisymmetry (QA) the  flux-function  $F(\psi)$ is given by
\begin{align}
 F(\psi)=\frac{G(\psi)}{\iota(\psi)},
\end{align}
where $2\pi G=\oint_\vartheta \B\cdot \bm{dr}$ is the poloidal current obtained by integrating $\B$ along a curve on the flux surface with a constant poloidal angle $\vartheta$, and $\iota(\psi)$ is the rotational transform.

A fundamental quantity is the quasisymmetry vector $\u$, which ensures that the field strength $B$ is constant along $\u$. Since $B$ is single-valued, $\u$ lines must be closed curves. In axisymmetry, $\u = R^2 \dl \phi$, where $(R,\phi,z)$ is the standard cylindrical coordinate system with $\phi$ denoting the symmetry angle. Thus, $\u$ is determined in axisymmetry purely from geometry and is decoupled from $\B$. However, $\u$ and $\B$ in QA are nontrivially coupled. The exact equations satisfied by $\u$ depend on whether we impose the so-called \textit{strong} or the \textit{weak} forms of QS \citep{burby2020,rodriguez2020a,constantin2021}. In the weak form, $\u$ satisfies
\begin{subequations}
    \begin{align}
    \B\times \u &= \dl \psi, \label{eq:Bxu}\\
    \uD B &=0, \label{eq:uDB}\\
    \dl \cdot \u &=0. \label{eq:div_u}
\end{align}
\label{eq:weak_QS}
\end{subequations}
In the strong form of QS, the first two equations for $\u$, \eqref{eq:Bxu} and \eqref{eq:uDB}, are unaltered, but the divergence-free condition \eqref{eq:div_u} is replaced by
\begin{align}
    \J\times \u +\dl (\u\cdot\B)=0.
    \label{eq:strong_QS_Jxu}
\end{align}
The equivalence of the various forms of QS with MHS force balance (under additional mild assumptions, such as a density of irrational flux surfaces) is well known \citep{rodriguez2020a,constantin2021,burby2020,rodrigGBC}. To keep the discussion self-contained, we have briefly discussed the equivalence of the various forms of QS in Appendix \ref{app:eqv_QS}. We focus on the strong form of QS for the rest of the work. We note that the divergence-free condition \eqref{eq:div_u} also holds in strong QS because strong QS always implies weak QS.

\subsection{Derivation of the quasisymmetric Grad-Shafranov system}
\label{sec:GS_picture}
The quasisymmetric generalized Grad Shafranov equation (GGSE) for $\psi$ was first derived in \citet{burby2020} assuming the strong form of QS and MHS. Due to the three-dimensional nature of $\psi$, additional conditions had to be imposed in order to satisfy the ideal MHS conditions. The system was expressed as a nonlinear overdetermined system for $(\u,\psi)$, expressed in the basis $(\u, \v, \w)$, where
\begin{align}
    \v \equiv \dl \times \u , \quad \w \equiv \v \times \u +\dl u^2.
    \label{eq:v_w_def}
\end{align}
A related quantity of importance is the \emph{strain tensor} $\mathsfbi{S}$ \citep{burby2024characterization}, defined such that
\begin{align}
    \mathsfbi{S}=\frac{1}{2}\lbr \nabla \u +{\nabla \u}^\top \rbr, \quad \w=\u\cdot \tnsr{S}.
\end{align}

The quantities $\tnsr{S}$ and $\w$ measure how much the QS vector $\u$ deviates from a Killing vector. If $\u$ is a Killing vector, $\tnsr{S}=0$ and $\w=0$ identically, and it can be shown that the flux surfaces are either axisymmetric or helically symmetric, both of which are two-dimensional. Therefore, we assume $\w\neq 0$.

The $(\u,\v,\w)$ system becomes degenerate near axisymmetry since $\w=0$ in axisymmetry. Therefore, we now develop a set of equations that involve the GGSE and the other constraints as an alternative to \citet{burby2020}. Details are provided in Appendix \ref{app:Derivation_GGSE}. 

We start with some basic results that follow directly from the strong form of QS. Dotting \eqref{eq:strong_QS_Jxu} with $\u$ and $\B$ and taking into account $\B\times \u=\dl \psi$ and $\J\cdot \dl \psi=0$ leads to $\BD (\u\cdot \B)=0$ and  $\uD (\u\cdot \B)=0$, which implies that $\u\cdot \B$ must be a flux function. Moreover, the conditions $\B\times \u=\dl \psi$ and  $\uD B=0$ imply that
\begin{align}
\u\cdot\B = F(\psi),
    \label{eq:udotB_eq}
\end{align}
where $F(\psi)$ is given by the 2-term form \eqref{eq:2term_QS}. Therefore, MHS together with the strong form of QS imply that the current $\J$ must be of the form
\begin{align}
    \J= - p' \u - F' \B.
    \label{eq:strong_form_J}
\end{align}

The expression for $\B$ in terms of $\u$ that follow from $\B\times \u =\dl \psi$ and $\u\cdot \B =F(\psi)$ is
\begin{align}
    \B = \frac{1}{u^2}(F(\psi)\u +\u \times \dl \psi).
    \label{eq:B_in_terms_of_u}
\end{align}
Thus, given $\u,\psi, F(\psi)$, we can obtain $\B$ from \eqref{eq:B_in_terms_of_u}. 

Before giving a rigorous derivation of the GGSE, let us give a quick derivation starting with the expression $\nabla^2 \psi \equiv \dl \cdot \dl \psi$. Using $\dl\psi=\B \times \u$, we find
\begin{align}
    \nabla^2  \psi = \J \cdot \u - \v \cdot \B.
\end{align}
The expressions for $ \J \cdot \u$ and $\v \cdot \B$ follow from the expression of $\J$ \eqref{eq:strong_form_J}, and the expression of $\B$ in terms of $\u$ \eqref{eq:B_in_terms_of_u}. We find $\J\cdot \u=-(p' u^2 +F F')$, whereas $\v\cdot \B$ can be expressed in terms of $\v \cdot \u$ and $\v \times \u \cdot \dl \psi$. We then obtain the following second-order elliptic equation for $\psi$,
\begin{align}
    \nabla^2  \psi -\frac{\u \times \v}{u^2}\cdot \dl \psi +\frac{\u\cdot \v}{u^2}F + FF' +u^2 p'=0.
    \label{eq:pre_GGS}
\end{align}
Equation \eqref{eq:pre_GGS} is solvable for $\psi$ \textit{only} when we know $\u$ and $\v$. Such is the case in axisymmetry, where $\u=R^2\dl\phi,\; \v=2\dl z$ in the standard cylindrical coordinates $(R,\phi,z)$, and \eqref{eq:pre_GGS} reduces to the standard axisymmetric Grad-Shafranov (GS) equation. However, in general geometry, $\u,\B ,\psi$ are not decoupled, which makes it unclear what is input or output. We show next that \eqref{eq:pre_GGS} can indeed be useful if it is supplemented by additional consistency conditions.

The MHS constraints on $\B$, namely $\dl\cdot \B=0$, $\J=\dl\times \B$, and $\J = -p' \u -F' \B$, need to be consistent with the expression for $\B$ in \eqref{eq:B_in_terms_of_u}. The current obtained by taking the curl of $\B$ from \eqref{eq:B_in_terms_of_u} takes the form
\begin{align}
    \J +(p' \u +F' \B) = \frac{1}{u^2}\J_w + \frac{1}{u^2} J_{\text{GGS}}\; \u.
    \label{eq:J_from_curl}
\end{align}
The expressions for $\J_w, J_{\text{GGS}}$ are provided in Appendix \ref{app:Derivation_GGSE}. The left side of \eqref{eq:J_from_curl} is zero from \eqref{eq:strong_form_J}. Equating the terms on the right to zero, we get the following two conditions
\begin{subequations}
\begin{align}
\nabla^2  \psi -\frac{\u \times \v}{u^2}\cdot \dl \psi +\frac{\u\cdot \v}{u^2}F + FF' +u^2 p'=0, \label{eq:GGSE}\\
\B\times \w = \uD \dl \psi - \dl \psi \cdot \dl \u. \label{eq:Bxw}
\end{align}
\label{eq:GGS_Bxw}
\end{subequations}
Equation \eqref{eq:GGSE} is the GGSE, and \eqref{eq:Bxw} leads to extra conditions that appear because of the intrinsic 3D nature of the fields. 

Both these equations were obtained by \citet{burby2020}, who then used the $(\u,\v,\w)$ basis to show that \eqref{eq:Bxw} leads to two nontrivial conditions. Since we are ultimately interested in near-axisymmetric systems with $\w\to 0$, we use the $(\u,\dl\psi, \B)$ basis instead, which remains non-degenerate as we approach axisymmetry. The $\u$ component of \eqref{eq:Bxw} is trivial as shown by  \citet{burby2020}. The $\dl \psi$ and the $\B$ components respectively lead to 
\begin{align}
    \uD B=0, \quad \dl\psi\cdot \tnsr{S}\cdot \B=0.
\end{align}

Unlike in the axisymmetric case, where $\w=0$, the divergence of $\B$ from \eqref{eq:B_in_terms_of_u} is not necessarily zero and leads to the constraint $\B \cdot \w = 0$, which reads
\begin{align}
    \BD u^2 -\v\cdot \dl \psi=0.
    \label{eq:Bdotw}
\end{align}
Alternatively, the $\dl\cdot \B=0$ condition given by \eqref{eq:Bdotw}, can also be written as $\u\cdot \tnsr{S}\cdot \B=0$.

To summarize, we start with the standard MHS equations \eqref{eq:ideal_MHS} with the QS constraint imposed through the strong QS (or equivalently weak QS and MHS)
\begin{align}
    \dl \cdot \u =0, \quad \u \cdot \dl B =0, \quad \B \times \u =\dl \psi, \quad \J =-(p' \u +F' \B),
\end{align}
together with the choice such that $\u\cdot \B=F(\psi)$, and $F\neq 0$ everywhere.

The quasisymmetric generalized Grad-Shafranov system is then given by
\begin{subequations}
    \begin{align}
     \uD \psi=0, \label{eq:uD_psi_is_0}\\
    \uD B=0, \label{eq:uD_B2_is_0}\\
   \dl \cdot \u =0, \label{eq:div_u_is_0}\\
    \BD u^2 -\v\cdot \dl \psi=0,
    \label{eq:Bdotw_is_0}\\
    \dl \psi \cdot \dl \u \cdot \B + \B \cdot \dl \u \cdot \dl \psi=0
    \label{eq:Jw_dot_del_psi_is_0}\\
    \nabla^2  \psi -\frac{\u \times \v}{u^2}\cdot \dl \psi +\frac{\u\cdot \v}{u^2}F + FF' +u^2 p'=0,\label{eq:GGSE_again}
\end{align}
\label{eq:Basic_QS_MHS_system}
\end{subequations}
where
\begin{align}
     \B = \frac{1}{u^2}(F(\psi)\u +\u \times \dl \psi) , \quad B^2=\frac{1}{u^2}\lbr F^2 +|\dl \psi|^2\rbr, \quad \v =\dl \times \u. \label{eq:B_form}
\end{align}

Equations \eqref{eq:uD_psi_is_0}, \eqref{eq:uD_B2_is_0} and \eqref{eq:div_u_is_0} determine components of $\u$ which must be divergence-free and orthogonal to $\dl B $ and $ \dl \psi$. In terms of $\u$ and $u^2$, we can compute $\B$ and $B^2$ from \eqref{eq:B_form}. Equation \eqref{eq:Bdotw_is_0} then enforces $\dl\cdot \B=0$. To impose $\J=\dl \times \B$ and $\J=-p'\u-F' \B$, we need to satisfy the $\uD B=0$ condition given by \eqref{eq:uD_B2_is_0}, the $\dl \psi \cdot \tnsr{S}\cdot B=0$ condition given by \eqref{eq:Jw_dot_del_psi_is_0}, and finally the GGSE equation \eqref{eq:GGSE_again}. The unknowns are the three components of $\u$ and $\psi$, $(u_1,u_2,u_3,\psi)$. There are six equations determining them in \eqref{eq:Basic_QS_MHS_system}, showing the overdetermined nature of the system.

In an axisymmetric system, we can describe the system \eqref{eq:Basic_QS_MHS_system} using the regular cylindrical coordinates $(R,\phi,z)$. The symmetry vector $\u=R^2\dl \phi$ is determined entirely by geometry and decouples from $\B$. We then have
\begin{align}
    \u = R^2\dl \phi, \;\; u=R,\;\;  \v=2 \dl z,\; \;\frac{\u \times \v}{u^2}= \frac{2\dl R}{R},\;\; \u\cdot \v=0.
    \label{eq:AS_u}
\end{align}
It follows that the only nontrivial equation in the GS system \eqref{eq:Basic_QS_MHS_system} is the GS equation \eqref{eq:GGSE_again}. From the identities \eqref{eq:AS_u} we can rewrite \eqref{eq:GGSE_again} in the standard form
\begin{align}
    \dlts \psi + F F' + R^2 p'=0, \quad \dlts \equiv R\del_R\lbr \frac{1}{R}\del_R\rbr+\del_z^2.
    \label{eq:GS_AS}
\end{align}

\section{Near-axisymmetric expansion (NASE)}
\label{sec:NASE}
We now expand about an axisymmetric solution using an expansion parameter $\epsilon$ that measures deviation from perfect axisymmetry. Since the lowest order is axisymmetric, using the standard cylindrical $(R,\phi,z)$ coordinates makes sense. Thus,
\begin{align}
    \u &= R^2 \dl \phi +\ep \u_1 +O(\ep^2 \u),\\
    \psi &= \psi_0(R,z)+\ep \psi_1(R,\phi,z)+O(\ep^2\psi)\nonumber,\\
    F(\psi) &= F(\psi_0)+\ep \psi_1 F'(\psi_0) +O(\ep^2 F).\nonumber
\end{align}
Note that the functions $\psi_1$ and $\psi_0$ and their gradients are comparable. The lowest order is given by \eqref{eq:AS_u} and \eqref{eq:GS_AS} for $\psi_0$, i.e.,
\begin{align}
\u_0 = R^2\dl \phi, \;\; u_0 =R,\;\;  \v_0=2 \dl z,\; \;\frac{\u_0 \times \v_0}{u_0^2}= \frac{2\dl R}{R},\;\; \u_0\cdot \v_0=0,\nonumber\\
\dlts \psi_0 + F F' + R^2 p'=0,\nonumber\\
\B_0 = F \dl \phi +\dl \phi \times \dl \psi_0, \quad \J_0 =-p' \u_0 -F' \B_0, \quad \tnsr{S}_0=0. \label{eq:O(1)}
\end{align}

We now proceed with the derivation of the first-order NASE system of equations \eqref{eq:O(ep))_cyl} (equivalently \eqref{eq:O(ep))_cyl_L}) from the basic quasisymmetric MHS system \eqref{eq:Basic_QS_MHS_system}. The details are provided in Appendix \ref{app:Derivation_NASE}. Employing the standard cylindrical coordinates with unit vectors given by  $(\hat{\bm{R}},\hat{\bm{\phi}},\hat{\bm{z}})$, the first-order vectors $\u_1, \v_1$ take the form
\begin{subequations}
    \begin{align}
    \u_1&= u_{1R} \Rh +  u_{1\phi}\hat{\phi} + u_{1z}\zh,\\
    \v_1&=\frac{\hat{\bm{R}}}{R}\lbr \del_\phi u_{1z}-\del_z (R u_{1\phi})\rbr +\hat{\bm{\phi}}\lbr \del_z u_{1R}-\del_R u_{1z}\rbr+\frac{\hat{\bm{z}}}{R}\lbr \del_R (R u_{1\phi})-\del_\phi u_{1R}\rbr.
\end{align}
\label{eq:u1_v1_components}
\end{subequations}

As shown in Appendix \ref{app:Derivation_NASE}, the first-order NASE system \eqref{eq:O(ep))_cyl} consists of six coupled first-order linear PDEs for the four scalars $((u_{1R},u_{1\phi},u_{1z},\psi_1(R,\phi,z))$. The toroidal geometry dictates the boundary conditions for the PDE system. In particular, $\psi_1$ must be periodic in the poloidal and toroidal angles $\theta,\phi$. Since the equations are linear and the coefficients are determined by the axisymmetric background, the $\phi$ dependence of all the first-order quantities can all be assumed to be of $\exp{(i N \phi)}$ form, as done in \citet{plunk2018,plunk2020_near_axisymmetry_MHD}. Therefore, instead of solving for $X_1(R,z,\phi)=X_{1c}(R,z)\cos{N\phi}+X_{1s}(R,z)\sin{N\phi}$, one could rewrite \eqref{eq:O(ep))_cyl} as a system of coupled equations for $X_{1c}(R,z), X_{1s}(R,z)$. Furthermore, because of the $\exp{(i N \phi)}$ dependence of the first-order quantities, we can make the identification
\begin{align}
    \del_\phi^2\leftrightarrow -N^2. \label{eq:N^2_id}
\end{align}

An alternative, more convenient form of the equations given by \eqref{eq:O(ep))_cyl_L} in Appendix \ref{app:Derivation_NASE}, can be obtained if we define $\Lambda_{1R},\Lambda_{1\phi},\Lambda_{1z},\Lambda_1$, such that
\begin{align}
u_{1R}=\del_\phi \Lambda_{1R}, \quad u_{1z}=\del_\phi \Lambda_{1z}, \quad u_{1\phi}=\del_\phi \Lambda_{1\phi},
\quad \Lambda_1\equiv \del_z \psi_0\; \Lambda_{1R}-\del_R\psi_0\; \Lambda_{1z}.
    \label{eq:Lambda_def}
\end{align}


\subsection{The structure of the NASE equations}
\label{sec:summary}
The first-order NASE system \eqref{eq:O(ep))_cyl_L} is linear but is overdetermined. Fortunately, the NASE system can be divided into ``on-surface" and ``off-surface" components. The former have angular (poloidal and toroidal) derivatives on a fixed $\psi_0$ surface, whereas the latter involve radial gradients (in $\dl\psi_0$ direction). The separation between on-surface and off-surface equations should allow for the possibility of a near-surface expansion \citep{sengupta2021NSE} at least locally near $\psi_0$.

A detailed analysis of the system \eqref{eq:O(ep))_cyl_L}, given in Appendix \ref{app:structure_preserving}, reveals the underlying structure of the near-axisymmetric quasisymmetric MHS equilibrium. The components of
$\u$,
$u_{1R}=\del_\phi \Lambda_{1R}, u_{1z}=\del_\phi \Lambda_{1z}, u_{1\phi}=\del_\phi \Lambda_{1\phi}$,
can be expressed in terms of only two functions, $\psi_1,\Lambda_1$, and their on-surface derivatives using
\begin{subequations}
    \begin{align}
    \Lambda_{1R}&=\frac{1}{|\dl\psi_0|^2}\lbr -\psi_1 \del_R\psi_0 +\Lambda_1\del_z \psi_0 \rbr,\label{eq:Lambda_1R_exp}\\
    \Lambda_{1z}&=-\frac{1}{|\dl\psi_0|^2}\lbr\psi_1 \del_z\psi_0 +\Lambda_1\del_R \psi_0 \rbr,\label{eq:Lambda_1z_exp}\\
    -N^2 \Lambda_{1\phi}&=
    \frac{R}{F^2}\{\psi_0,\del_\phi\Lambda_1\}+ \hat{\cA}_1 \del_\phi\psi_1+ \hat{\cA}_2\del_\phi\Lambda_1.\label{eq:Lambda_1phi_exp}
\end{align}
\label{eq:Lambda_i_so}
\end{subequations}
The various coefficients such as $\cA_1, \cA_2, \cA_3, \cA_4, \hat{\cA}_1, \hat{\cA}_2, C_\Psi, C_\psi, C_\Lambda, C_\phi$ that appear in the NASE equations are given in Appendix \ref{app:structure_preserving}. Here, and elsewhere, we use the notation $\{f,g\}_{(z,R)}$ to denote the usual Poisson bracket of variables $f$ and $g$ with respect to $z$ and $R$ respectively. The subscript $(z,R)$ will usually be omitted for brevity. The bracket $[\cL_1, \cL_2]\equiv \cL_1 \cL_2-\cL_2\cL_1$ denotes the usual commutator of two operators $\cL_1$ and $\cL_2$. 

The functions $(\psi_1,\Lambda_1)$ satisfy the following linear PDE that only has second-order (on-surface) angular derivatives
\begin{align}
    &\lbr \frac{1}{R^2}\del_\phi +\frac{2F}{R^2}\frac{\del_z \psi_0}{|\dl\psi_0|^2}\rbr\Lambda_1-\frac{2F \del_R \psi_0}{R^2|\dl\psi_0|^2}\psi_1\label{eq:on_surf_psi_Lambda}\\
    &=\frac{1}{N^2}\lbr \frac{2F}{R}\del_\phi+\{\psi_0,\;\; \} \rbr\lbr \frac{1}{F^2}\{\psi_0,\del_\phi\Lambda_1\}+ \frac{1}{R}\lbr \hat{\cA}_1 \del_\phi\psi_1+ \hat{\cA}_2\del_\phi\Lambda_1\rbr\rbr,\nonumber
\end{align}
where $\{\psi_0,\;\; \}$ denotes an operator acting to the right, such that $\{\psi_0,\;\; \}X  = \{\psi_0, X\}$.
An alternate form of \eqref{eq:on_surf_psi_Lambda}, which will be useful later for the WKB analysis, is given in \eqref{eq:on_surf_psi_Lambda_alt_bal}.

Once \eqref{eq:on_surf_psi_Lambda} (alternatively  \eqref{eq:on_surf_psi_Lambda_alt_bal}) is solved subject to standard periodicity in the poloidal angle, and $\exp{(i N \phi)}$ in the toroidal angle, the vector $\u_1$ can be computed. With the quasisymmetry vector $\u_1$ and the first-order flux surface $\psi_1$, we can then compute the first-order quasisymmetric $\B_1$ using
\begin{align}
    \B_1 = -2 \frac{\u_0 \cdot \u_1}{R^2}\B_0 +\frac{1}{R^2}\lbr F \u_1 +\psi_1 F' \u_0 + \u_0\times \dl \psi_1 +\u_1 \times \dl \psi_0\rbr.
\end{align}
To extend the solution to a volume, we use the following off-surface or the ``radial" gradients of the quantities $\psi_1$ and $ \Lambda_1$
\begin{subequations}
    \begin{align}
    &\dl\psi_0\cdot \dl \psi_1=\frac{(RB_0)^2}{F^2}\lbr \{\psi_0,\Lambda_1\}-\frac{|\dl\psi_0|^2}{R}\lbr (\cA_1+\cA_3)\psi_1 + (\cA_2+\cA_4)\Lambda_1\rbr \rbr,\label{eq:grad_psi1_dot_grad_psi0}\\
&\dl\psi_0\cdot\dl\Lambda_1=\lbr\frac{|\dl\psi_0|}{R B_0}\rbr^2\frac{R F}{N}\left( \lbr \dlts +\cC_\Psi\rbr \frac{1}{N}\del_\phi\psi_1 - \frac{N F}{R|\dl\psi_0|^2}\{\psi_0,\psi_1\}  \right.\nonumber\\
& \qquad\qquad \qquad \qquad\qquad \left. +\frac{N F}{F^2}\B_0\cdot\dl\psi_1+ N\left( \cC_\psi \psi_1 +\cC_\Lambda\Lambda_1+\cC_\phi \Lambda_{1\phi}\right)\right).\label{eq:grad_psi1_dot_grad_Lambda_1T}
\end{align}
\label{eq:radial_grads}
\end{subequations}
The radial derivative of $\Lambda_{1\phi}$ follows from
\begin{align}
     F \dl\psi_0\cdot \dl \lbr \frac{\Lambda_{1\phi}}{R}\rbr =- \frac{1}{R}\dl\psi_0\cdot \dl \Lambda_1 +\B_0\cdot \dl \psi_1 +\frac{\Lambda_1}{R}\lbr \dlts\psi_0 +\frac{1}{R}\del_R \psi_0\rbr\label{eq:grad_psi1_dot_grad_Lambda_1phi}.
\end{align}
We have interpreted the GGSE \eqref{eq:GGS_cyl_L} here as an equation determining the radial gradient of $\Lambda_1$.  

The compatibility condition that leads to the magnetic shear is obtained from the surface average of
\begin{align}
    &\left[\dl\psi_0\cdot \dl, \B_0\cdot \dl\right](R\Lambda_{1\phi})+\B_0\cdot \dl \lbr \del_\phi\psi_1 +\dl\psi_0\cdot \dl (R\Lambda_{1\phi}) \rbr \nonumber\\
    &+\frac{1}{R}\del_\phi\lbr \dlts\psi_0  \Lambda_1+F' |\dl\psi_0|^2\Lambda_{1\phi}\rbr  +\dl\psi_0\cdot \dl \lbr \frac{2}{R}\lbr F \Lambda_{1R}-\del_z\psi_0 \Lambda_{1\phi}\rbr \rbr=0.
    \label{eq:towards_shear}
\end{align}
Magnetic shear in near-axis expansions can also be shown to arise from similar compatibility conditions \citep{rodriguez2021weak,sengupta2024NAE_finite_beta}.

Summarizing, we find that the first-order NASE system of equations for compact QAS devices can be split into a set of on-surface and off-surface equations for $\psi_1,\Lambda_1$. Every other quantity of interest,  namely, $(\Lambda_{1R},\Lambda_{1z},\Lambda_{1\phi})$, can be obtained from $(\psi_1,\Lambda_1)$ from \eqref{eq:Lambda_i_so}.  The second-order PDE \eqref{eq:on_surf_psi_Lambda} determines $(\psi_1,\Lambda_1)$ on a surface of constant $\psi_0$. The off-surface equations that provide the radial ($\psi_0$) gradients for the $(\psi_1,\Lambda_1)$ system are given by \eqref{eq:radial_grads}. The GGSE appears as a radial gradient equation for $\Lambda_1$. There is an additional equation for the radial derivative of $\Lambda_{1\phi}$ \eqref{eq:grad_psi1_dot_grad_Lambda_1phi}. Since $\Lambda_{1\phi}$ can be determined from $(\psi_1,\Lambda_1)$, \eqref{eq:grad_psi1_dot_grad_Lambda_1phi} leads to a consistency condition \eqref{eq:towards_shear}, involving the self-consistent magnetic shear.

Given the NASE system's overdetermination, a volumetric solution is unlikely to exist. A more mathematical treatment dealing with the overdetermined system's consistencies and higher orders of the NASE is left for the future.

\subsection{Ridges in near-axisymmetric compact devices}
\label{sec:Ridges_LargeN}

Consistent with the nonexistence result in the vacuum limit \citep{plunk2018} and the overall increase in QS error with a decrease in aspect ratio \citep{henneberg2024compact,Schuett_henneberg2024compactQA,plunk2020_near_axisymmetry_MHD,buller2024family}, it is clear that even if a solution can be found, it will be highly constrained. Therefore, instead of trying to find general solutions to the NASE system, our primary goal in this work is to obtain a characterization of a critical feature of most compact QA stellarators: the presence of sharp ridges on the outermost flux surface area. We aim to explore and understand the interplay between geometry, QS, and MHS force balance that leads to the formation of these sharp structures.

The ridges play an important role in divertor design \citep{bader2019} because the magnetic fields tend to align with the ridges, imparting them an almost separatrix-like behavior. The ridges with characteristic sharp variations on the flux surface typically occur near the minimum of $|\dl\psi|$. On the other hand, a magnetic axis (denoting X-points in this context) will have $\dl\psi=0$. Consistent with Paper I, we shall define ridges as structures along which $|\dl\psi|$ is at its minimum but not necessarily zero everywhere.  

Expanding around the minimum of $|\dl\psi|$ allows us to obtain the following closed set of linear equations that we expect to describe the ridges. 
\begin{subequations}
\begin{align}
& R \B_0\cdot \dl \Lambda_1=\frac{F}{R}\del_\phi \Lambda_1+\frac{|\dl\psi_0|^2}{R}\lbr (\cA_1+\cA_3)\psi_1 + (\cA_2+\cA_4)\Lambda_1\rbr,\label{eq:grad_psi1_dot_grad_psi0_bal}\\
&(R \B_0\cdot \dl)^2\Lambda_1+F^2 R \B_0\cdot \dl\left( \frac{1}{R}\lbr \hat{\cA}_1 \psi_1+ \hat{\cA}_2\Lambda_1\rbr\right)\label{eq:on_surf_psi_Lambda_alt_bal}\\
    &+\lbr\frac{F}{R}\rbr^2 F\del_\phi \lbr  \lbr \hat{\cA}_1 -2\frac{\del_R \psi_0}{|\dl\psi_0|^2}\rbr \psi_1+ \lbr\hat{\cA}_2+2\frac{\del_z \psi_0}{|\dl\psi_0|^2}\lbr 1+\frac{|\dl\psi|^2}{2F^2}\rbr\rbr\Lambda_1\rbr=0.\nonumber
\end{align}
    \label{eq:2_bal_eqns}
\end{subequations}
\noindent The equations reflect that ridges locally minimize $|\dl\psi|^2$ and hence $\dl\psi_1\cdot \dl\psi_0\approx 0$ in \eqref{eq:grad_psi1_dot_grad_psi0}, and the on-surface equation \eqref{eq:on_surf_psi_Lambda_alt_bal} that follows from $\B$ being divergence-free. 

In the following, we will discuss the physical consequences of \eqref{eq:2_bal_eqns} by exploiting two subsidiary limits. The first assumes a large number of field periods $N$ with periodic boundary conditions imposed on the solution. The second assumes an eikonal solution with a ballooning-like boundary condition.

\section{The large $N$ subsidiary ordering assumptions}
\label{sec:LargeN_ordering_assumptions}
To capture the ridge-like features analytically in the NASE system, we now examine the limit of a large number of field periods, $N$. In principle, a ridge can occur for low $N>2$, and only the poloidal variation needs to be large. In either case, a large on-surface gradient implies that the local surface curvature will be extremized, as shown in \citet{bader2019}. The assumption of large $N$ simplifies the analysis of the PDEs considerably and offers excellent insight into the nature of the ridges. In this Section, we look at how such sharp structures naturally emerge from the NASE system in the large $N$ limit and investigate what constraints they impose on the axisymmetric background. We shall show in Section \ref{sec:Numerical} that our theory offers insight also for low $N$ such as $N=3$.

Invoking toroidal periodicity, we can assume that the nonsymmetric terms are of the form $X_1\sim f_1(R,z)\exp{(i N \phi)}$. We now carry out a subsidiary expansion in large $N$, which implies large variations in the toroidal direction. We assume that $R_0|\dl X_0|/X_0\sim O(1), (\del_\phi X_1)/X_1\sim O(N)$, where $R_0$ is a typical major radius and $X_0(R,z), X_1(R,\phi,z)$ are the lowest-order (axisymmetric) and first-order (nonsymmetric) quantities. Since QS is not a symmetry of the geometry, the assumption of strong toroidal $\phi$ gradients of the first-order perturbations implies that poloidal $\vartheta$ variations must also be large. 

Therefore, we consider the first-order NASE system as a system of overdetermined linear PDEs describing the fast variations of the perturbations on top of a slowly varying axisymmetric background. Denoting normalization with $R_0$ by a bar, our subsidiary ordering is formally
\begin{align}
&\frac{|\overline{\dl} X_0(R,z)|}{X_0}= O(1), \quad \frac{1}{X_1}(\del_\phi, \del_\vartheta, \overline{\del}_R, \overline{\del}_z)X_1(R,z,\phi)= O(N). \label{eq:large_N_subsidiary_main}
\end{align}

At this point, it is convenient to introduce the normalized operators 
\begin{align}
    \del_{N \phi} \equiv \frac{1}{N}\del_\phi, \;\; \text{such that} \;\;  \del_{N\phi}^2\leftrightarrow-1,\;\; \dlts_N \equiv \frac{1}{N^2}\dlts, \;\;\text{and}\;\; \{\;\;,\;\;\}_N \equiv \frac{1}{N}\{\;\;,\;\;\}.
\end{align}

So far, we have not discussed the ordering for the radial or $\psi$ variations. We discuss this in the next Section since it is dictated by the conditions that lead to the formation of ridges.

\subsection{Analysis of the radial variations imposed by the ridges}
\label{eq:ridges_radial_analysis}
As discussed in Paper I, we expect $\dl\psi_0\cdot \dl \psi_1$ to be at its minimum near ridges. This imposes constraints on the various geometrical and physical quantities that are associated with it through MHS force balance and QS. Therefore, \eqref{eq:grad_psi1_dot_grad_psi0} is a key equation since it determines $\dl\psi_0\cdot \dl\psi_1$. Although in an infinitely sharp ridge the minimum of $|\nabla \psi|$ is exactly zero as discussed in Paper I, it is unclear what the minimum is in the general case. We now discuss this further in the large $N$ limit.

On substituting the large $N$ subsidiary expansion,
\begin{align}
\Lambda_1&=\Lambda_1^{(0)} + N^{-1} \Lambda_1^{(1)}+O(N^{-2}\Lambda_1), \quad \Lambda_{1\phi}=\Lambda_{1\phi}^{(0)} + N^{-1} \Lambda_{1\phi}^{(1)}+ O(N^{-2}\Lambda_{1\phi}), 
   \nonumber \\
    \psi_1 &= \psi_1^{(0)} + N^{-1} \psi_1^{(1)}+O(N^{-2}\psi_1), \label{eq:LargeN_subs}
\end{align}
into \eqref{eq:grad_psi1_dot_grad_psi0} we find
\begin{align}
   &-N\{\psi_0,\Lambda_1^{(0)}\}_N -\{\psi_0,\Lambda_1^{(1)}\}_N +O(N^{-1})\nonumber\\
   &+\frac{F^2}{(RB_0)^2} \dl\psi_0\cdot \dl \psi_1^{(0)} +\frac{|\dl\psi_0|^2}{R}\lbr (\cA_1+\cA_3)\psi_1^{(0)} + (\cA_2+\cA_4)\Lambda_1^{(0)}\rbr=0.
\end{align}
Since the ridges become increasingly sharp as $N$ increases, as shown in Paper I for vacuum fields, we expect $\dl\psi_1\cdot \dl\psi_0$ to vanish to lowest order in $1/N$. The condition $(\dl\psi_1\cdot \dl\psi_0)/|\nabla \psi_0|^2\sim O(N^{-1})$, allows for finite $N$ corrections and leads to
\begin{align}
    \Lambda_1^{(0)}=0,\quad \{\psi_0,\Lambda_1^{(1)}\}_N- \frac{|\dl\psi_0|^2}{R} (\cA_1+\cA_3)\psi_1^{(0)} =0.
    \label{eq:Lambda_IT_N}
\end{align}
Consequently, it follows from \eqref{eq:Lambda_i_so} that
\begin{subequations}
    \begin{align}
    \Lambda_{1R}^{(0)}&= -\psi_1^{(0)} \frac{\del_R\psi_0}{|\dl\psi_0|^2} ,\quad
    \Lambda_{1z}^{(0)}= -\psi_1^{(0)} \frac{\del_z\psi_0}{|\dl\psi_0|^2}, \quad \Lambda_{1\phi}^{(0)}=0\label{eq:Lambda_1R_1z_1phi_exp_N}\\
    -\Lambda_{1\phi}^{(1)}&=
    \frac{R}{F^2}\{\psi_0,\del_{N\phi}\Lambda_1^{(1)}\}_N+ \hat{\cA}_1 \del_{N\phi}\psi_1^{(0)}.\label{eq:Lambda_1phi_exp_N}
\end{align}
\label{eq:Lambda_i_struct_N}
\end{subequations}

Another key equation is the one determining the radial gradient of $\Lambda_{1\phi}$ \eqref{eq:grad_psi1_dot_grad_Lambda_1phi}, i.e.,
\begin{align}
    \frac{F}{N} \dl\psi_0\cdot \dl \lbr \frac{\Lambda_{1\phi}^{(1)}}{R}\rbr =- \frac{1}{R N}\dl\psi_0\cdot \dl \Lambda_1^{(1)} +\B_0\cdot \dl \psi_1^{(0)} +O(N^{-1}).
    \label{eq:BDpsi1_N}
\end{align}
Several important observations follow from \eqref{eq:BDpsi1_N}. First, we note that since the perturbed quantities have large gradients in both poloidal and toroidal angles, both $\B_0\cdot \dl\psi_1 $ and $\B_1 \cdot \dl \psi_0$ formally scale with $N$ separately but they must cancel each other since $\BD\psi=0$ order by order. 
Next, we order $\bm{n}\cdot \overline{\dl} \sim O(1)$, where $\bm{n}=\nabla \psi_0/|\nabla \psi_0|$ is the unit vector normal to the lowest order flux surface. We get $\B_0\cdot \dl\psi_1^{(0)}=0$ from \eqref{eq:BDpsi1_N}, which implies that $\psi_1^{(0)}$ should not vary along the lowest order magnetic field. Unless the rotational transform $\iota_0$ is rational, this condition implies that $\psi_1=\psi_1(\psi_0)$, i.e., $\psi_1$ must be an axisymmetric function of $\psi_0$ to lowest order, which would mean $\psi_1=0$ to lowest order since we are interested only in nonsymmetric first-order variations. For rational rotational transform, $\psi_1=\psi_1(\psi_0,\alpha_0)$, where $\alpha_0$ denotes a closed field line. Thus, constant $\psi_1$ contours and hence the ridges follow the lowest order closed field line. The more general ordering $\bm{n}\cdot \overline{\dl} \sim O(N)$ or higher will permit $\B_0\cdot \dl\psi_1^{(0)}$ to be nonzero and hence the rotational transform can be irrational. Note that $\bm{n}\cdot \overline{\dl} \sim O(N^2)$ is permitted since $\B_0\cdot\dl \psi_1$ can be formally scale linearly with $N$.

Thus, we obtain the following ordering that supports large-$N$ ridge-like solutions
\begin{align}
    &\bm{n}\cdot \overline{\dl}\psi_1 \sim O(N^{-1}\psi_1), \quad \frac{1}{N X_1}\bm{n}\cdot \overline{\dl} X_1\sim \frac{1}{\psi_1}\B_0\cdot \overline{\dl} \psi_1, \quad \bm{n}\equiv \frac{\nabla \psi_0}{|\nabla \psi_0|},\quad \overline{\dl}\equiv R_0 \dl, \nonumber \\
   &\frac{1}{N X_1}\bm{n} \cdot \overline{\dl} X_1 \sim O(N) \;\; \text{(general)},\quad\frac{1}{X_1}\bm{n} \cdot \overline{\dl} X_1 \sim O(1)\;\; \text{(rational $\iota_0$)}
   \label{eq:ridge_conditions}. 
\end{align}

\subsection{On-surface characteristics that determine the ridges}
\label{sec:ridges_on_surf_analysis}
We have seen in the previous Section that the lowest order system can be described by the $\lbr \Lambda_1^{(1)},\psi_1^{(1)}\rbr$ system where
\begin{align}
\{\psi_0,\Lambda_1^{(1)}\}_N = \frac{|\dl\psi_0|^2}{R} (\cA_1+\cA_3)\psi_1^{(0)}
    \label{eq:Lambda_1T_1psi_N1}
\end{align}
Substituting \eqref{eq:Lambda_1T_1psi_N1} into the expression for $\Lambda_{1\phi}^{(1)}$ in \eqref{eq:Lambda_1phi_exp_N} we get
\begin{align}
\Lambda_{1\phi}^{(1)} =-\hat{\cA}_{1\phi}\del_{N\phi}\psi_1^{(0)}, \quad \hat{\cA}_{1\phi}\equiv \hat{\cA}_1 + \frac{|\dl\psi_0|^2}{F^2}\lbr\cA_1+\cA_3\rbr=-\cA_3.
    \label{eq:Lambda_1T_1phi_psi1_N1}
\end{align}

The on-surface PDE \eqref{eq:on_surf_psi_Lambda} takes the form
\begin{align}
    \frac{1}{R^2}\del_{N\phi}\Lambda_1^{(1)}=\frac{2F \del_R \psi_0}{R^2|\dl\psi_0|^2}\psi_1^{(0)}+\lbr \frac{2F}{R}\del_{N\phi}+\{\psi_0,\;\; \}_N \rbr\lbr \frac{1}{R}\hat{\cA}_{1\phi}\del_{N\phi}\psi_1^{(0)}\rbr.
    \label{eq:on_surf_psi_Lambda_N1}
\end{align}
Eliminating $\Lambda_1^{(1)}$ between \eqref{eq:Lambda_1T_1psi_N1} and \eqref{eq:on_surf_psi_Lambda_N1} we get a single second-order hyperbolic PDE for $\psi_1^{(0)}$. 
Formally defining an angle $\theta$ such that
\begin{align}
    (F/R)\del_{N\theta}\equiv \{\psi_0,\;\}_N,
    \label{eq:formal_Ntheta}
\end{align}
we can rewrite the PDE for $\psi_1^{(0)}$ as a hyperbolic Laplace equation \citep{forsyth_v4_pt6} that appears in the theory of conjugate nets \citep{eisenhart1923transformations}
\begin{align}
    \del_{N\theta}\lbr R\del_{N\theta}\lbr\frac{\hat{\cA}_{1\phi}}{R}\psi_1^{(0)}\rbr+2\lbr \hat{\cA}_{1\phi}-\frac{\del_R\psi_0}{|\dl\psi_0|^2}\rbr\del_{N\phi}\psi_1^{(0)}\rbr\nonumber\\
    =\frac{|\dl\psi_0|^2}{F^2}\lbr\cA_1+\cA_3\rbr\psi_1^{(0)}.
    \label{eq:on_surf_theta_phi_psi}
\end{align}
Ideally, one would like to solve the on-surface equation \eqref{eq:on_surf_theta_phi_psi} using the standard WKB formalism. We will address this later in Section \ref{sec:bal_ridges}. For now, we will be content with a local solution, assuming that the coefficients determined by the axisymmetric background do not change on the fast scales. Then we can obtain a local solution assuming a Fourier basis i.e., $\psi_1^{(0)}\sim \exp{i N(k_\theta \theta \pm \phi)}$. We find that $k_\theta$ must satisfy the quadratic equation
\begin{align}
    &\cC_1 k_\theta^2 \pm 2\cC_2 k_\theta +\cC_0=0, \label{eq:k_theta_quad}\\
    &\cC_1\equiv \hat{\cA}_{1\phi}=-\cA_3, \;\; \cC_2= \lbr \hat{\cA}_{1\phi}-\frac{\del_R\psi_0}{|\dl\psi_0|^2}\rbr, \;\; \cC_0= \frac{|\dl\psi_0|^2}{F^2}\lbr\cA_1+\cA_3\rbr. \nonumber
\end{align}
The reality condition for the solution of the quadratic equation for $k_\theta$ that ensures periodicity in $\theta$ is 
\begin{align}
\cC_1\neq 0, \quad \cC_2^2-\cC_1\cC_0\geq 0.
    \label{eq:realtor_kc}
\end{align}
In the degenerate cases, with $\cC_2\neq 0$, but $(\cC_0=0,\cC_1\neq 0)$ and $(\cC_0\neq 0,\cC_1=0)$ we obtain $(k_\theta=0,\pm 2\cC_2/\cC_1)$ and $(k_\theta=\pm\cC_0/(2\cC_2))$ respectively. Since $\cC_0,\cC_1$ are functions of $(R,z)$, these degenerate cases can occur only around certain points and are not general.

Provided the reality condition \eqref{eq:realtor_kc} holds, the ridge-like solutions are characterized by perturbations that have a characteristic pitch determined by $k_\theta$ satisfying the ``dispersion relation" \eqref{eq:k_theta_quad}. If the reality condition is not satisfied, the only solution is the trivial solution $\psi_1^{(0)}=0$, which implies no nonsymmetric perturbations. Since the background axisymmetric equilibrium determines the pitch through the coefficients $\cC_i$'s, the reality condition imposes stringent restrictions on the axisymmetric background, which will be the subject of our next Section.

\subsection{Conditions for the existence of ridge-like solutions}
\label{eq:existence_of_ridges}
From the expressions of various $\cA_i$'s given in \eqref{eq:AS_coeffs}, we find that the reality condition \eqref{eq:realtor_kc} is a condition on $\cA_0$ since
\begin{align}
     \cC_0= \frac{R}{F^2}\lbr \lbr \nabla^2 \psi_0-\frac{\del_R\psi_0}{R} \rbr-\cA_0 \lbr 2-\frac{|\dl\psi_0|^2}{(RB_0)^2}\rbr\rbr \label{eq:cCj_expr}\\
    \cC_1=- \frac{R}{(RB_0)^2}\cA_0+\frac{\del_R \psi_0}{|\dl\psi_0|^2}, \quad \cC_2=-\frac{R}{(RB_0)^2}\cA_0.\nonumber
\end{align}
Defining the following quantities,
\begin{align}
    \cA_{00}\equiv \delta_J +\delta_0, \quad \cA_{01}^2\equiv  \delta_J^2 +\delta_0\delta_R, \quad\delta_0\equiv \delta_R(1+\varepsilon_p^2),\\
    \vep_p^2\delta_R \equiv \frac{\del_R\psi_0}{R}, \quad 2\delta_J \equiv \nabla^2 \psi_0 -2\frac{\del_R\psi_0}{R}=\dlts\psi_0, \quad \varepsilon_p \equiv \frac{|\dl\psi_0|}{F},
\end{align}
where $\delta_J$ is a measure of currents beyond vacuum, $\delta_R$ and $\delta_0$ are measures of toroidal curvature effects, $\varepsilon_p$ is a measure of the ratio of the poloidal to the toroidal flux. With these definitions, we find that
\begin{align}
\cC_1\neq 0\quad \Leftrightarrow \quad \cA_0 \neq \delta_0.
    \label{eq:nonzerocC1}
\end{align}
Further, we can rewrite the reality condition as $\cD_F\geq 0$, where
\begin{align}
    \cD_F\equiv (\cC_2^2-\cC_0\cC_1)\lbr \frac{(RB_0)F}{R}\rbr^2=-\lbr (\cA_0 - \cA_{00})^2 -\cA_{01}^2 \rbr\geq 0 .
    \label{eq:realtor_alt}
\end{align}
It follows $\cD_F\geq 0$ that real ridge-like solutions exist only in the range
\begin{align}
    \cA_{00}-|\cA_{01}|\leq \cA_0 \leq  \cA_{00}+|\cA_{01}|, \quad \text{with}\quad \cA_0 \neq \delta_0.
    \label{eq:Realtor_sol}
\end{align}
We now look at some of the physically interesting limits of \eqref{eq:Realtor_sol}. First, since both the aspect ratio $(a/R_0)$ and the rotational transform $\iota$ of a QA device, particularly vacuum QA, are typically smaller than unity, $\varepsilon_p=(a/R_0)\iota\leq 1$. For QA close to vacuum axisymmetry, $\varepsilon_p\ll 1$, even for a compact device as $\iota\ll 1$. Next, we note that $\delta_J=-(R^2p'+FF')$ from lowest-order force-balance \eqref{eq:O(1)}. Therefore, in the vacuum limit, $\delta_J=0,\; \cA_{00}=\delta_0,\;\cA_{01}^2=\delta_0\delta_R$ and for small $\vep_p$, \eqref{eq:Realtor_sol} implies
\begin{align}
     \frac{1}{2}\vep_p^2\delta_R\leq \cA_0 \leq  \vep_p^2\delta_R\lbr \frac{1}{\vep_p^2}+\frac{3}{2} +O(\vep_p^2)\rbr.
    \label{eq:Realtor_low_beta_range}
\end{align}
For low-beta systems, typical of stellarators, $\delta_J$ is small, and \eqref{eq:Realtor_low_beta_range} is only slightly modified through the replacement of $\cA_0$ by $\cA_0-\delta_J$. The only solution outside of this range is the trivial axisymmetric solution. As discussed in \cite{nikulsin_sengupta2024GS}, a device with volumetric QS can typically support large pressure gradients provided the aspect ratio is large. Therefore, we shall now assume that $\delta_J$ is small in the following analysis. We shall discuss the case of sizeable pressure gradient later in Section \ref{sec:bal_ridges}.

\subsection{The quasisymmetric constraints on the Gaussian curvature}
The constraint on $\cA_0$ in \eqref{eq:Realtor_low_beta_range} translates directly into a constraint on the Gaussian curvature as we show in the following. The special case of $\cA_0=0$, which from \eqref{eq:AS_coeffs} is equivalent to $\dl\psi_0\cdot \dl |\dl\psi_0|^2=0$, is a parabolic PDE called the \textit{Aronsson equation} 
\citep{aronsson_crandell2004tour,aronsson1968partial}.
The Gaussian curvature $\cK$ is intimately related to $\cA_0$ \citep{aronsson1968partial}. In particular, for an axisymmetric geometry, as shown in Appendix \ref{app:K}, the expressions for $\cK$ and $\cA_0$ are
\begin{subequations}
    \begin{align}
    \cK &= \lbr \frac{\del_R\psi_0}{R}\rbr\frac{\lbr (\del_z\psi_0)^2\del^2_R\psi_0 +(\del_R\psi_0)^2\del^2_z\psi_0 -2 (\del_R\psi_0)(\del_z \psi_0)\del_R\del_z\psi_0\rbr}{\lbr(\del_R\psi_0)^2 +(\del_z\psi_0)^2\rbr^2},\label{eq:cK_gen_exp}\\
    \cA_0 &=\frac{\lbr (\del_R\psi_0)^2\del^2_R\psi_0 +(\del_z\psi_0)^2\del^2_z\psi_0 +2 (\del_R\psi_0)(\del_z \psi_0)\del_R\del_z\psi_0\rbr}{\lbr(\del_R\psi_0)^2 +(\del_z\psi_0)^2\rbr},
    \label{eq:cA_0_gen_exp}\\
    F^2\cK &=\delta_R \lbr 2\delta_J +\delta_R \vep_p^2-\cA_0\rbr.
    \label{eq:cK_cA0_reln}
\end{align}
\label{eq:cA_0_and_cK}
\end{subequations}
From the range of $\cA_0$ \eqref{eq:Realtor_low_beta_range} we find that the Gaussian curvature must be within the range,
\begin{align}
    \frac{1}{2}\lbr \vep_p\frac{\delta_R}{F}\rbr^2\geq \cK \geq  -\lbr \vep_p\frac{\delta_R}{F}\rbr^2\lbr \frac{1}{\vep_p^2}+\frac{1}{2} +O(\vep_p^2)\rbr.
    \label{eq:cK_low_beta_range}
\end{align}
Therefore, in the vicinity of the 3D ridge, the Gaussian curvature normalized by $R_0$ is mostly negative and covers a range
\begin{align}
 \frac{1}{2}\lbr\frac{R_0\delta_R}{F}\rbr^2 \vep_p^2     \geq R_0^2\cK \geq    -\lbr\frac{R_0\delta_R}{F}\rbr^2
    \label{eq:cK_low_beta_range_real}
\end{align}
We note that for $\delta_R=0$, equivalently $\del_R\psi_0=0$ but $|\dl\psi_0|\neq 0$,  the allowed range collapses to $\cK = 0$. This corresponds to two circles on the top and bottom of the axisymmetric torus. If additionally $\dl\psi_0=0$, the circles degenerate to X-points for the axisymmetric field. 

We further notice that ridges strongly prefer to be on the negative Gaussian curvature side of the torus for $\delta_R\neq 0$. Since the Gaussian curvature is both positive and negative on a torus, the perturbation $\psi_1$ must vanish in the region $\cK > (\vep_p^2\delta_R^2)/(2F^2)$.  Typically, Gaussian curvature assumes its maximum positive value on the outboard side of a torus, which implies that $\psi_1$ must be zero on the outboard side. 

Further geometrical constraints on the axisymmetric background can be obtained by observing that $\psi_0$ must satisfy an equation of the form 
\begin{align}
    (\del_R\psi_0)^2\lbr \del_R^2\psi_0-\hat{\cA}_0 \rbr+(\del_z\psi_0)^2\lbr \del_z^2\psi_0-\hat{\cA}_0 \rbr +2(\del_R\psi_0)(\del_z\psi_0) \del_R\del_z \psi_0 =0,
    \label{eq:quad_A0}
\end{align}
where $\hat{\cA}_0$ is a particular value of $\cA_0$ within the range \eqref{eq:Realtor_low_beta_range}. For real solutions to \eqref{eq:quad_A0} viewed as a quadratic equation for $(\del_z \psi_0)/(\del_R \psi_0)$, we must have
\begin{align}
    \lbr \del_R \del_z \psi_0\rbr^2 -\lbr \del_R^2\psi_0-\hat{\cA}_0 \rbr\lbr \del_z^2\psi_0-\hat{\cA}_0 \rbr\geq 0.
\end{align}

\subsection{The GGSE in the large $N$ limit}
\label{sec:GGSE_largeN}
We have obtained the characteristics of the ridge-like solutions on a surface in Section \ref{sec:ridges_on_surf_analysis}. In this Section, we look at the solution of the GGSE \eqref{eq:GGSE_comp} in the large $N$ limit under the ridge conditions \eqref{eq:ridge_conditions}. As discussed in Section \ref{sec:summary}, the GGSE yields the off-surface gradient of $\Lambda_1$. The ridge-like conditions \eqref{eq:ridge_conditions} differentiate two cases. The first case is the general case with irrational ($\iota_0$), and the radial derivative of $\Lambda_1$ scales as $N^2$. The second case assumes a rational $\iota_0$ and the radial gradient to be $O(1)$. We shall discuss these cases now to demonstrate the internal consistency of the various assumptions we have made that led to the ridge-like conditions \eqref{eq:ridge_conditions}.

The normalized large $N$ GGSE for $\psi_1^{(0)}$ reads $(N R_0)^2 T_{-2}  +(N R_0) T_{-1} \sim \psi_1^{(0)}$, where
\begin{align}
    T_{-2}\equiv \lbr \dlts_N\del_{N\phi}\psi_1^{(0)}-\frac{F}{R|\dl\psi_0|^2}\{\psi_0,\psi_1^{(0)}\}_N\rbr,\label{eq:lowest_order_GGSE}\\
    T_{-1} \equiv \frac{1}{F}\B_0\cdot\dl\psi_1^{(0)} + \cC_\psi \psi_1^{(0)}-\lbr\frac{R B_0}{|\dl\psi_0|}\rbr^2\frac{1}{N R F} \dl\psi_0\cdot\dl\Lambda_1^{(1)}.\nonumber
\end{align}
We now analyze the consequences of the ridge conditions \eqref{eq:ridge_conditions}. Firstly, we can safely ignore derivatives of the background in comparison to derivatives of the perturbed quantities, consistent with our assumptions in \eqref{sec:LargeN_ordering_assumptions}. Secondly, we note that for irrational $\iota_0$, both $N^2 T_{-2}$ and $N T_{-1}$ scale with $N^2$ since the ridge conditions imply that $\B_0\cdot \dl \psi_1\sim (1/N)\dl\psi_0\cdot \dl \Lambda_1$ scale with $N$. Thus, the GGSE determines the radial gradient of $\Lambda_1^{(1)}$ by balancing $T_{-2}$ and $T_{-1}$, whereas the $O(\psi_1^{(0)})$ term on the right is a higher order term. 

Next, we utilize the ridge condition $(\dl\psi_0 \cdot \dl \psi_1)/|\nabla \psi_0|^2 \sim O(1/N)$ to rewrite $\del_z \psi_1^{(0)}$ in terms of $\del_R \psi_1^{(0)}$. Commuting the derivatives past the background quantities we obtain
\begin{align}
&-\del_z \psi_1^{(0)}=\frac{\del_R \psi_0}{\del_z \psi_0} \del_R\psi_1^{(0)}, \;  \del_z^2\psi_1^{(0)}\approx \lbr \frac{\del_R \psi_0}{\del_z \psi_0}\rbr^2 \del_R^2\psi_1^{(0)} \Rightarrow \dlts \psi_1^{(0)}\approx  \lbr \frac{|\dl \psi_0|}{\del_z \psi_0}\rbr^2 \del_R^2\psi_1^{(0)} \nonumber\\
&\left\{\psi_0,\psi_1^{(0)} \right\} = \lbr \frac{|\dl \psi_0|}{\del_z \psi_0}\rbr^2 (\del_z \psi_0)\del_R\psi_1^{(0)}.  \label{eq:del_z_replaced}
\end{align}
The lowest-order GGSE \eqref{eq:lowest_order_GGSE} then takes the simple form 
\begin{align}
  \lbr \frac{|\dl \psi_0|}{\del_z \psi_0}\rbr^2  \del_R\lbr \del_R + \frac{F}{R}\frac{\del_z\psi_0}{|\dl\psi_0|^2}\del_{\phi} \rbr \del_{N\phi}\psi_1^{(0)}+N T_{-1}\sim \psi_1^{(0)}.
  \label{eq:lowest_GGSE_2_terms}
\end{align}
Upon using the expression
\begin{align}
 \B_0\cdot \dl \psi_1^{(0)} =  \frac{1}{R}\frac{|\dl \psi_0|^2}{\del_z \psi_0}\lbr \del_R +\frac{F}{R} \frac{\del_z\psi_0}{|\dl\psi_0|^2} \del_\phi\rbr \psi_1^{(0)},
 \label{eq:B0Dpsi10_exp}
\end{align}
\eqref{eq:lowest_GGSE_2_terms} becomes
\begin{align}
 \B_0\cdot \dl \lbr \frac{R}{\del_z \psi_0}\del_R \del_{N\phi}\psi_1^{(0)} \rbr +N T_{-1}\sim \psi_1^{(0)}.
 \label{eq:lowest_GGS_BD_form}
\end{align}
Ignoring the higher order term on the right, the lowest order GGSE reduces to 
\begin{align}
    \lbr\frac{R B_0}{|\dl\psi_0|}\rbr^2\frac{1}{N R } \dl\psi_0\cdot\dl\Lambda_1^{(1)}= \B_0\cdot \dl \lbr \psi_1^{(0)}+\frac{F R/N}{\del_z \psi_0}\del_R \del_{N\phi}\psi_1^{(0)} \rbr,
\end{align}
which is an equation that determines the radial derivative of $\Lambda_1^{(0)}$ for irrational $\iota_0$.

For the case of rational $\iota_0$, a similar analysis can be carried out to show that
\begin{align}
    \B_0\cdot \dl\psi_1^{(0)}=0, \quad \cC_\psi=0.
\end{align}
Therefore, the ridge conditions \eqref{eq:ridge_conditions} are self-consistent with the GGSE equation when $\cC_\psi=0$. Moreover, for rational nonzero $\iota$, a nonzero solution of $\B_0\cdot \dl\psi_1^{(0)}=0$ can also be obtained in a straightforward manner using the method of characteristics. It follows that for $\psi_1^{(0)}\neq 0$, we must have $\del_\varphi\psi_1^{(0)}=0 $, where
\begin{align}
\varphi\equiv \phi -\int \frac{dR}{R}\frac{F}{|\dl\psi_0|} \del_z \psi_0.
    \label{eq:varphi_def}
\end{align}
Thus, in the large $N$ limit, constant $\varphi$ curves denote sharp ridges, which are X-points in the $(R,\phi,z)$ space when $\iota$ is rational. Since the closed field lines cover the whole torus, the rational ridges (X-points) appear on both the inboard $(\cK<0)$ and the outboard ($\cK>0$) sides. 

In the case of $\iota_0=0$, a possible solution of $\cC_\psi=0$, defined in \eqref{eq:AS_coeffs}, is given by
\begin{align}
    \del_z\psi_0=0, \quad \B_0\cdot \dl |\dl\psi_0|^2=0.
    \label{eq:cpsi_0_condn}
\end{align}
One possibility is $\del_R\psi_0=0,\del_z \psi_0=0, \cK=0$. Another possibility is that $\del_R \psi_0 \neq 0$ but $\del_z \psi_0, \del_{R}\del_z\psi_0, \del_z^2 \psi_0$ all vanish simultaneously. In both these cases, $\varphi\approx \phi$, i.e., the flux surfaces are almost axisymmetric. We shall address this case later when we look at the $\iota\approx 0$ case.

\section{Finite pressure and WKB formalism for ridges with arbitrary $N$}\label{sec:bal_ridges}


We now present an alternative approach based on aperiodic boundary conditions, which are suitable for describing localized ridges. Utilizing the linearity of the fundamental equations that describe the first-order deviations from axisymmetry, we can use the well-known eikonal or ballooning-like formalism from the theory of ideal MHD stability \citep{dewar1983ballooning} to study the ridges. 

It is convenient to use the coordinate system $(\alpha_0,\psi_0,\vartheta)$ coordinate system where, $\vartheta$ is a suitably defined poloidal angle and $\alpha_0$ is the lowest-order field-line label such that
\begin{align}
    \B_0 =\dl\alpha_0 \times \dl \psi_0, \quad \dl\alpha_0\times \dl\psi_0\cdot \dl \vartheta = \cJ^{-1}, \quad \B_0\cdot \dl = \cJ^{-1}\del_\vartheta.
\end{align}
From axisymmetry, we know that $\del_\phi \alpha_0$ is unity because we choose $\psi$ as the poloidal flux. Thus, assuming a large gradient in the $\alpha_0$ direction compared to the field-line direction,
\begin{align}
    \at{\del_\phi}{(z,R)}= \del_{\alpha_0} \to i k_\alpha \gg 1.
\end{align}
The large gradient naturally leads to a WKB form for the solutions, $(\psi_1,\Lambda_1)\sim (\psi_{1a},\Lambda_{1a})\exp{(i k_\alpha S)} $, where the amplitudes $\psi_{1a},\Lambda_{1a}$ and the eikonal $S$, are assumed to vary slowly along the field line. Note that for large $N$, $k_\alpha$ can be replaced by $N$. Just as in standard ballooning theory, we shall assume that $\vartheta$ is defined in the covering space such that $-\infty <\vartheta<\infty$. The periodic boundary conditions on the perturbed quantities are relaxed. We only require that they not grow as $\vartheta \to \pm \infty$.

Similar to the large-$N$ subsidiary, expanding the coupled $(\Lambda_1,\psi_1)$ system \eqref{eq:2_bal_eqns} in inverse powers of $k_\alpha$, we find that
\begin{align}
\Lambda_1= \frac{1}{k_\alpha}\Lambda_1^{(1)} +O(k_\alpha^{-2}\Lambda_1), \quad \psi_{1}=\psi_{1}^{(0)}+\frac{1}{k_\alpha}\psi_{1}^{(1)} +O(k_\alpha^{-2}\psi_1).
    \label{eq:k_alpha_exp}
\end{align}
The lowest order, \eqref{eq:2_bal_eqns} then takes the form
\begin{subequations}
    \begin{align}
    \lbr \frac{1}{k_\alpha}\del_\parallel -i\rbr \lbr \frac{\Lambda_1^{(1)}}{F}\rbr= \cC_0 \psi_{1}^{(0)}, \quad \del_\parallel\equiv \frac{R}{F/R}\B_0\cdot\dl,
    \label{eq:lambda11_psi10_reln1}\\
  \frac{1}{k_\alpha^2}\del_\parallel^2   \lbr \frac{\Lambda_1^{(1)}}{F} \rbr+ \lbr \cC_1-\cC_0\rbr\frac{1}{k_\alpha}\del_\parallel  \psi_{1}^{(0)} +i\lbr 2\cC_2 -(\cC_0+\cC_1)\rbr\psi_{1}^{(0)}=0.
    \label{eq:lambda11_psi10_reln2}
\end{align}
 \label{eq:lowest_amplitude_relns}
\end{subequations}
The quantities $\cC_j$, where $j=0,1,2$,  previously appeared in the large-$N$ case and are given in \eqref{eq:k_theta_quad} and \eqref{eq:cCj_expr}. 

The system \eqref{eq:lowest_amplitude_relns} can be reduced to the following linear second-order ODE for $\Lambda_1^{(1)}$ along the lowest order $\B_0$
\begin{align}
 \cC_1\frac{1}{k_\alpha^2}\del_\parallel^2   \Lambda_1^{(1)}+ 2i\lbr \cC_1-\cC_0\rbr\frac{1}{k_\alpha}\del_\parallel  \Lambda_1^{(1)} +\lbr 2\cC_2 -(\cC_0+\cC_1)\rbr\Lambda_1^{(1)}=0.
    \label{eq:bal_main_eqn}
\end{align}

First, assuming $\cC_1\neq 0$, we can bring \eqref{eq:bal_main_eqn} to the Schr\"odinger equation form by defining
\begin{align}
    \Lambda_1^{(1)}=\chi_{1}\exp{\lbr -i k_\alpha S_\mfc\rbr}, \quad S_\mfc = \int \frac{F\cJ}{R^2} \lbr \frac{\cC_2-\cC_1}{\cC_1} \rbr d\vartheta,
\end{align}
such that
\begin{align}
    \frac{1}{k_\alpha^2}\del_\parallel^2 \chi_{1} + \frac{1}{\cC_1^2}\cD_B \chi_{1}=0, \quad \cD_B =  \cC_2^2-\cC_1\cC_0 .
\end{align}
Thus, the ``effective potential" is given by $\cD_B/\cC_1^2$. For the entire range where $\cC_1\neq 0$, the effective potential is periodic and we can, in principle, use Floquet-Bloch theory \citep{magnus2013hill_eq} to show that suitable eigenfunctions can always be constructed such that the zero boundary condition at infinity is satisfied. However, it is simpler to exploit the large $k_\alpha$ parameter and construct the WKB solution for $\chi_1$ as follows
\begin{align}
    \chi_1 =\frac{\chi_{1a}}{(\sqrt{\cD_B}/\cC_1)^{1/2}}\exp{\lbr\pm i k_\alpha S_\cD\rbr}, \quad S_\cD = \int  \frac{F\cJ}{R^2} \frac{\sqrt{\cD_B}}{\cC_1 } d\vartheta.
\end{align}
From the WKB solution for $\chi_1$ we get
\begin{align}
    \Lambda_1\approx \chi_{1a}e^{i k_\alpha S}\sqrt{\frac{\cC_1}{\sqrt{\cD_B}}},\; \psi_1 \approx \frac{-i \chi_{1a}e^{i k_\alpha S}}{\cC_0 }\sqrt{\frac{\cC_1}{\sqrt{\cD_B}}},\;
    S= \int \frac{F\cJ}{ R^2} d\vartheta \frac{\lbr \cC_1-\cC_2 \pm \sqrt{\cD_B} \rbr}{\cC_1 }. \label{eq:approx_WKB}
\end{align}
we readily find that the boundary condition of non-growing solutions implies that 
\begin{align}
    \cD_B= \cC_2^2-\cC_1\cC_0 \geq 0,
    \label{eq:bal_cD}
\end{align}
which is the same condition $\cD_F\geq 0$ that we obtained using periodic boundary conditions. 

The WKB solution \eqref{eq:approx_WKB} is inadequate near $\cC_1=0$ and $\cD_B=0$. From the definitions of $\cC_j$ given in \eqref{eq:cCj_expr}, we see that $\cC_1$ and $\cD_B$ do not vanish simultaneously unless $\del_R \psi_0=\cA_0=0$. We now examine these two cases, where we need to go beyond the standard WKB.

The breakdown of the WKB solution near $\cD_B\approx 0$ is the standard caustic condition \citep{berry1976waves_and_Thom} since $\cD_B=0$ represents ``turning points" for the Schr\"odinger equation. This issue is standard and can be easily resolved by connecting the WKB theory with Airy functions \citep{berry1976waves_and_Thom,dewar1983ballooning}. The behavior of wave functions near turning points is oscillatory on one side ($\cD_B>0$) and exponential on the other side ($\cD_B<0$). Therefore, our nonsymmetric solutions approach zero exponentially once they cross the $\cD_B=0$ point. 

Next, we analyze the case of $\cC_1=0,\cC_0\neq 0, \cC_2\neq 0$ such that $\cD_B\neq 0$, which reduces \eqref{eq:bal_main_eqn} to a first-order ODE. It follows that
\begin{align}
\Lambda_1^{(1)}=\Lambda_{1a}e^{ i k_\alpha S_{\cC_1}},\;\; \psi_1^{(0)}= \frac{-i\Lambda_1^{(1)}}{ F\cC_0}\lbr\frac{\cC_2}{\cC_0}-\frac{3}{2}\rbr, \;\; S_{\cC_1} = \int \frac{F\cJ}{ R^2} \lbr \frac{\cC_2}{\cC_0}-\frac{1}{2}\rbr d\vartheta.
    \label{eq:C1zero}
\end{align}
Since $\cC_1$ is zero only at finitely many points and not zero everywhere, the solution \eqref{eq:C1zero} is not generic. However, the behavior of the second-order ODE \eqref{eq:bal_main_eqn} in the vicinity of points with vanishing $\cC_1$ needs further exploration. The WKB solution has zero amplitude near the zeros of $\cC_1$.

Since $\cC_1$ is periodic, there are infinitely many points where $\cC_1=0$. The general theory of such ODEs \citep{coddington1956theoryODE} is far too complicated for the present analysis. Instead, we will focus on a single interval $-\vartheta_- \leq \vartheta \leq \vartheta_+$ such that $\cC_1$ has simple zeros at the upper and lower bounds $\vartheta_\pm$ of the interval. Since $\cC_1\to 0$, in general, imply that $\cC_0,\cC_2$ are finite, we can expand the ODE \eqref{eq:bal_main_eqn} near a zero of $\cC_1$ to obtain an approximate ODE of the form
\begin{align}
    \lbr x \frac{d^2}{dx^2} -2 i x_0 \frac{d}{dx} + x_2\rbr \Lambda_1^{(1)}=0,
    \label{eq:regular_sing}
\end{align}
where $x$ denotes the suitably normalized $(\vartheta-\vartheta_\pm)$ coordinate, and $x_0,x_2$ are the normalized values of the coefficients of \eqref{eq:bal_main_eqn} near $\vartheta_\pm$. Since \eqref{eq:regular_sing} is an ODE with a regular singularity \citep{coddington1956theoryODE}, the local behavior of the solutions near $x=0$ is of the form $x^\nu$. The indices turn out to be $\nu=0$ and $ 1+2i x_0$. Since we are looking for solutions that approach zero at large $\vartheta$, by choosing the second index, we can obtain a solution that approaches zero near $x=0$, which matches the WKB solution in the limit $\cC_1\to 0$. Outside of the range $-\vartheta_- \leq \vartheta \leq \vartheta_+$ we have the axisymmetric solution with $\psi_1=0$. Effectively, the zeros of $\cC_1$ act as infinite barriers and help localize the ridge-like solutions within the range of nonzero $\cC_1$.


\section{Numerical verification}
\label{sec:Numerical}
We have presented an analytical theory of near-axisymmetric ridges. We have derived the ridge conditions \eqref{eq:ridge_conditions} that enable us to obtain a closed set of equations \eqref{eq:2_bal_eqns} that determine the perturbed quantities $\Lambda_1,\psi_1$. We discussed approximate localized solutions of the ridge equations with periodic and ballooning-like boundary conditions making necessary subsidiary expansions for analytical tractability. 

In Section \ref{sec:Nikita_numerics} we solve the ridge equations \eqref{eq:2_bal_eqns} directly numerically without making any further assumptions to validate our analytical predictions of localized ridges. In Section \ref{sec:Rogerio_numerics} we present several optimized compact QA configurations using the SIMSOPT code \citep{landreman2021simsopt} with number of field periods, $n_{fp}$, varying between two and seven. We then test how our linear NASE theory performs against the fully nonlinear numerically optimized compact QA equilibria.

\subsection{Numerical analysis of the NASE system}
\label{sec:Nikita_numerics}

We will now numerically solve the ridge equations, which were derived in previous sections, on a single flux surface. We will use the analytical tokamak equilibrium from \citet{pataki_cerfon_2013fastGS} as the zeroth order axisymmetric term $\psi_0$. This solution can be written as follows:
\begin{equation}
    \psi_0(R,z) = C\left[R^4/8 + d_1 + d_2 R^2 + d_3(R^4 - 4R^2 z^2)\right],\label{eq:psi0}
\end{equation}
where $C$ is a constant with dimensionality $[B]/[L]^2$, which governs the strength of the poloidal field, and the $d_i$ coefficients are found by imposing the boundary condition $\psi_0 = 0$ at the plasma edge. We refer the reader to \citet{pataki_cerfon_2013fastGS} for details on how the boundary condition is imposed. In the examples shown below, $C = 1~\mathrm{T/m^2}$, $d_1 = 0.07538503~\mathrm{m^4}$, $d_2 = -0.20629496~\mathrm{m^2}$ and $d_3 = -0.03143371$.

\begin{figure}
    \centering
    \includegraphics[width=0.32\linewidth]{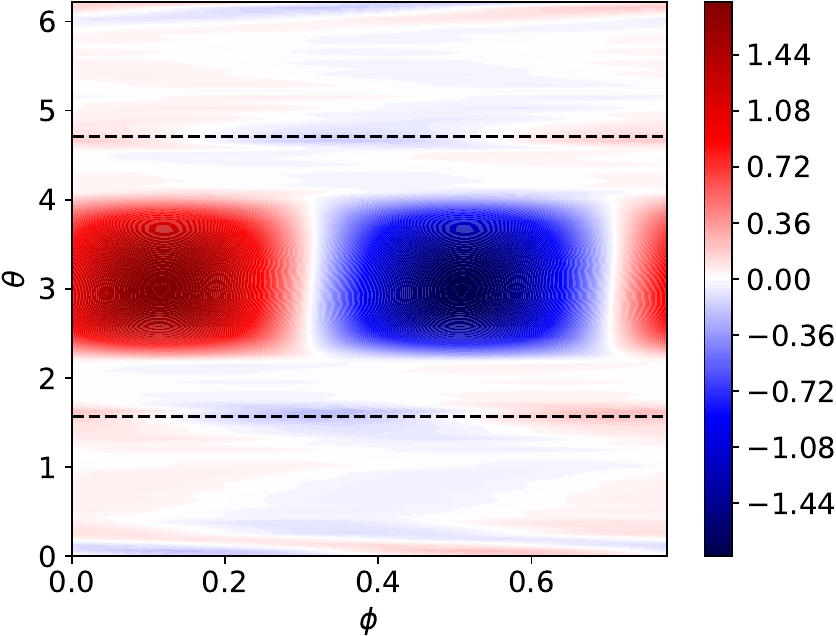} \includegraphics[width=0.32\linewidth]{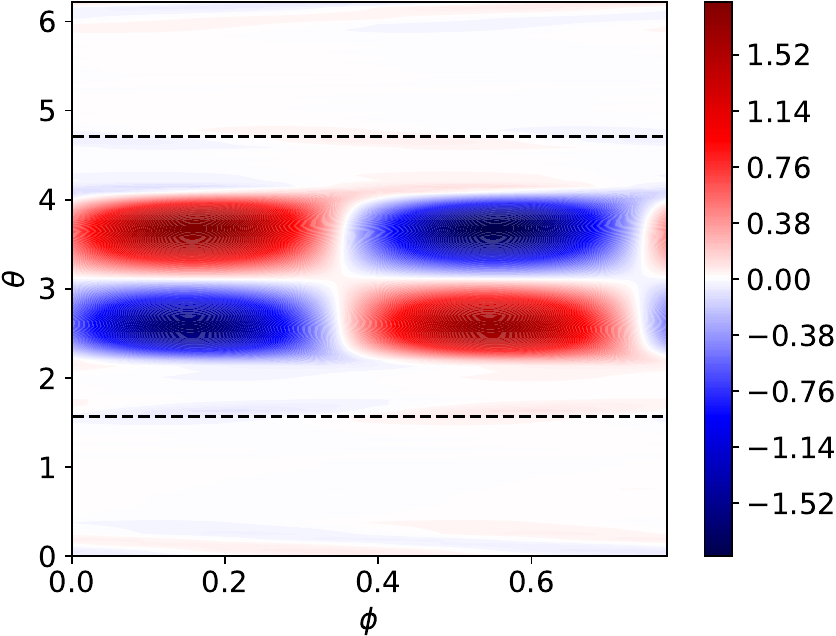} \includegraphics[width=0.32\linewidth]{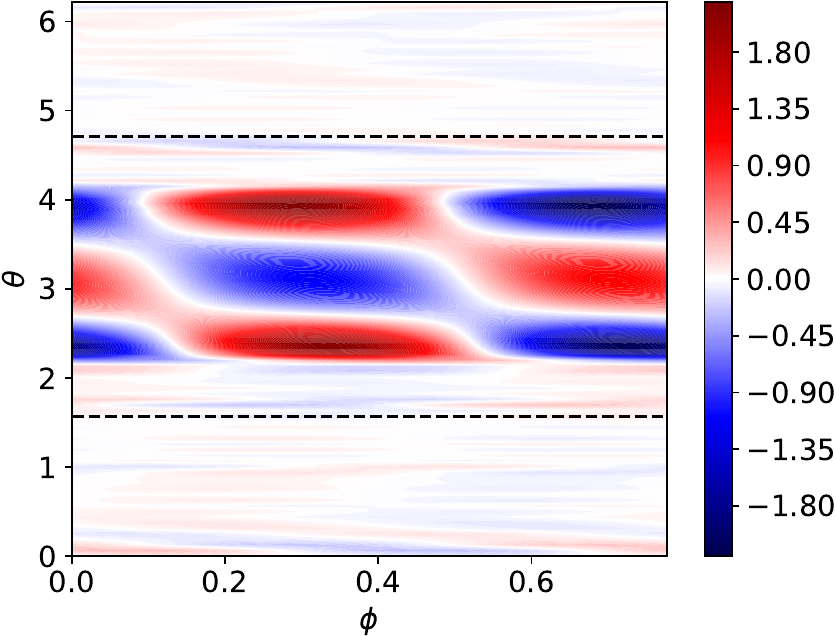} \\
    \includegraphics[width=0.32\linewidth]{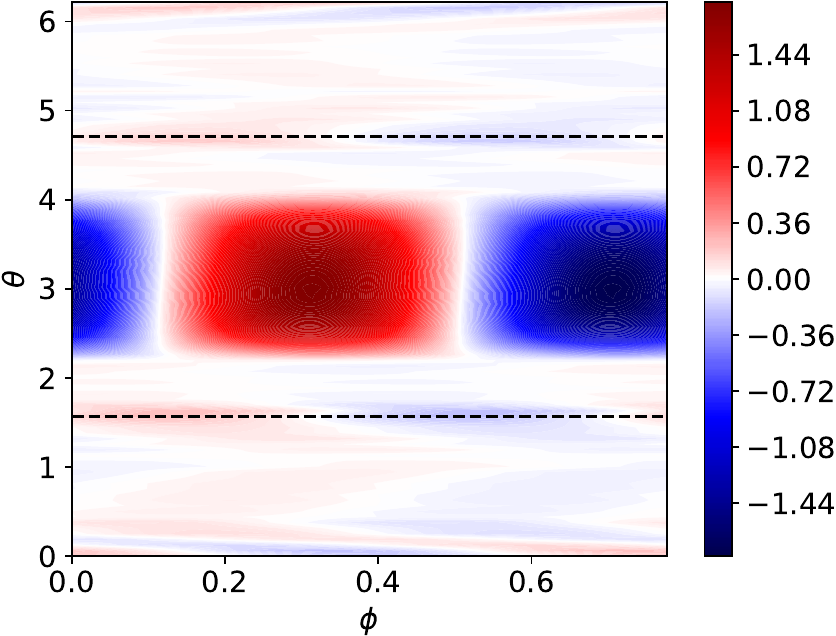} \includegraphics[width=0.32\linewidth]{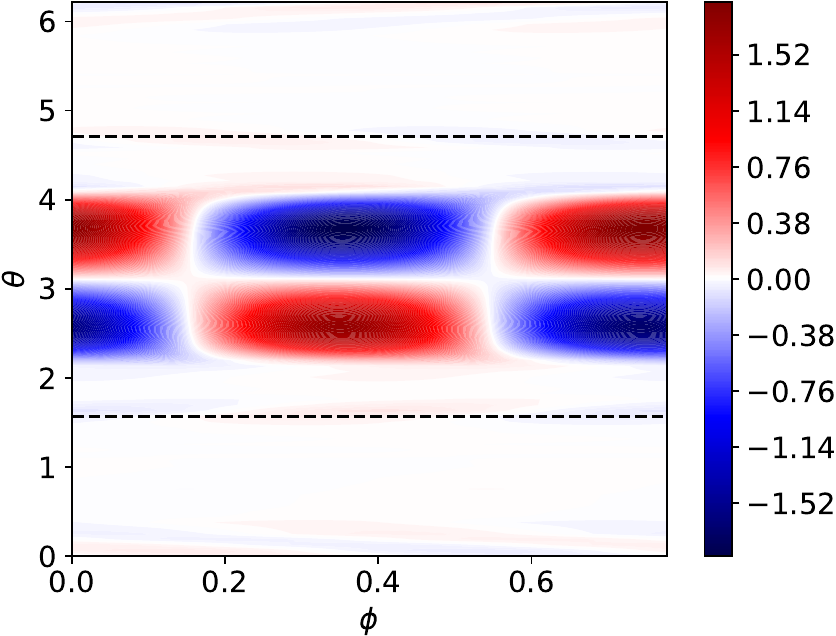} \includegraphics[width=0.32\linewidth]{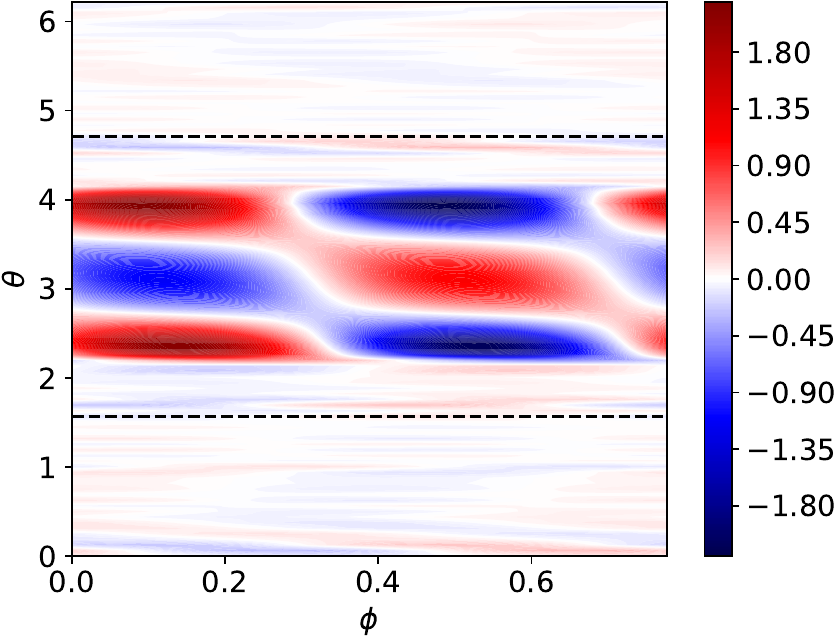}
    \caption{Several numerical solutions to equations \eqref{eq:2_bal_eqns} for $\psi_1$ with $N=8$ and $m=0,...,54$, plotted over one toroidal field period. The normalized residual $|\mathbf{Ax}|/\|\mathbf{A}\|_2$, where $\|\cdot\|_2$ is the spectral norm, is quite low, ranging from $\sim 6.0\times10^{-4}$ (left column) to $\sim 1.4\times10^{-3}$ (right column). The dashed lines correspond to the lines of zero Gaussian curvature at $\theta = \pi/2, 3\pi/2$, which is the boundary between the inboard and outboard sides.}
    \label{fig:nase_fourier}
\end{figure}

We will first solve equations \eqref{eq:2_bal_eqns} in a periodic domain using a Fourier-Galerkin method. Just as in Section \ref{sec:LargeN_ordering_assumptions}, we will consider only one toroidal Fourier mode, $\{\Lambda_1(\theta,\phi),\psi_1(\theta,\phi)\} = \Re\{\widetilde{\Lambda}_1(\theta),\widetilde{\psi}_1(\theta)\}e^{iN\phi}$, where $\theta = \arctan[z/(R-R_0)]$ is the poloidal angle ($R_0$ is the $R$-coordinate of the magnetic axis), and the radial coordinate $\psi_0$ is fixed to be the outermost flux surface. The number of poloidal modes can be varied, and is increased until convergence is achieved. In order to solve equations \eqref{eq:2_bal_eqns}, which have a trivial solution, we will need to treat this as an eigenvalue problem. After discretizing equations \eqref{eq:2_bal_eqns} via the Fourier-Galerkin method, we obtain the matrix equation $\mathbf{Ax} = 0$, where $\mathbf{A}$ is a known real (if we separate the real and imaginary parts of the equations and variables and treat them as independent) matrix and $\mathbf{x}$ is the solution vector in the Fourier basis. Since numerical solutions are necessarily approximate, and since equations \eqref{eq:2_bal_eqns} are anyway not accurate beyond $O(\epsilon)$, we should not demand that the matrix equation be satisfied exactly, but rather try to minimize $|\mathbf{Ax}|^2$ subject to $\mathbf{x}\neq 0$:
\begin{equation}
    \min_\mathbf{x} |\mathbf{Ax}|^2,\text{ s. t. }~~ |\mathbf{x}| = \mathcal{N} \implies \mathcal{L} = |\mathbf{Ax}|^2 + \lambda(|\mathbf{x}|^2 - \mathcal{N}^2),
\end{equation}
where $\mathcal{N}\neq 0$ is some normalization parameter and $\lambda$ is a Lagrange multiplier. Setting the derivative of the Lagrangian to zero gives
\begin{equation}
    \mathbf{A}^\top\mathbf{Ax} + \lambda\mathbf{x} = 0.\label{eq:eigvalform}
\end{equation}
Thus, we can use the existing SciPy \citep{virtanen2020scipy} eigenvalue solver for Hermitian matrices and keep only those eigenvectors for which the residuals $|\mathbf{Ax}|$ are sufficiently small. Figure \ref{fig:nase_fourier} shows $\psi_1$ from some numerically well-behaved such solutions using $\psi_0$ from \eqref{eq:psi0} with $N=8$, a poloidal resolution of $m=0,...,54$, where $m$ is the poloidal mode number. Note that, (1), the solutions are localized to the inboard side of the device, centered around $\theta = \pi$, and (2), the solutions look like a Fourier basis that was compressed to the region of localization, meaning that a superposition of these solutions can produce a ridge-like structure.

\begin{figure}
    \centering
    \includegraphics[width=0.32\linewidth]{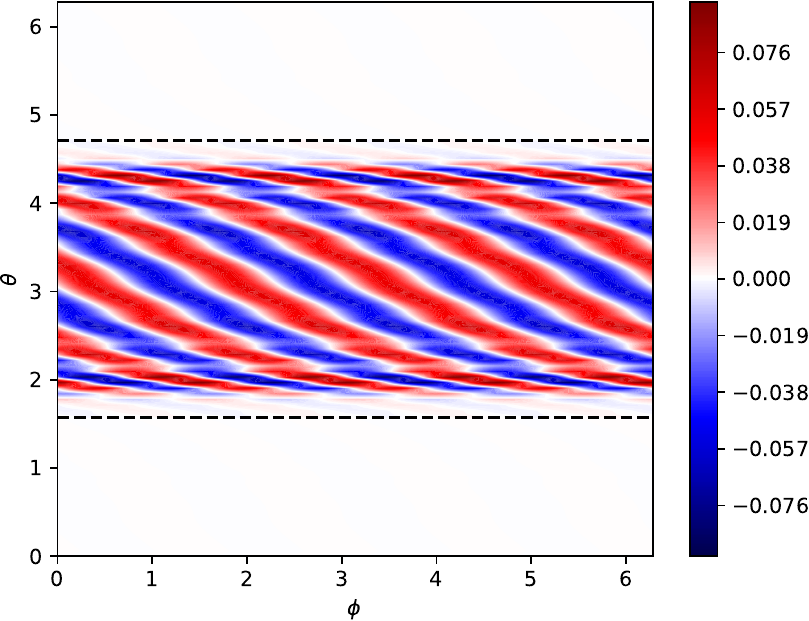} \includegraphics[width=0.32\linewidth]{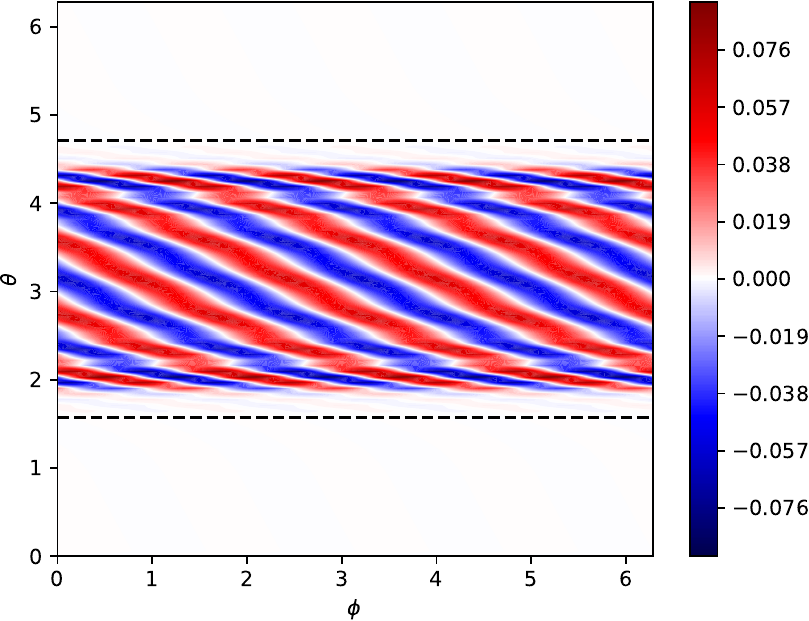} \includegraphics[width=0.32\linewidth]{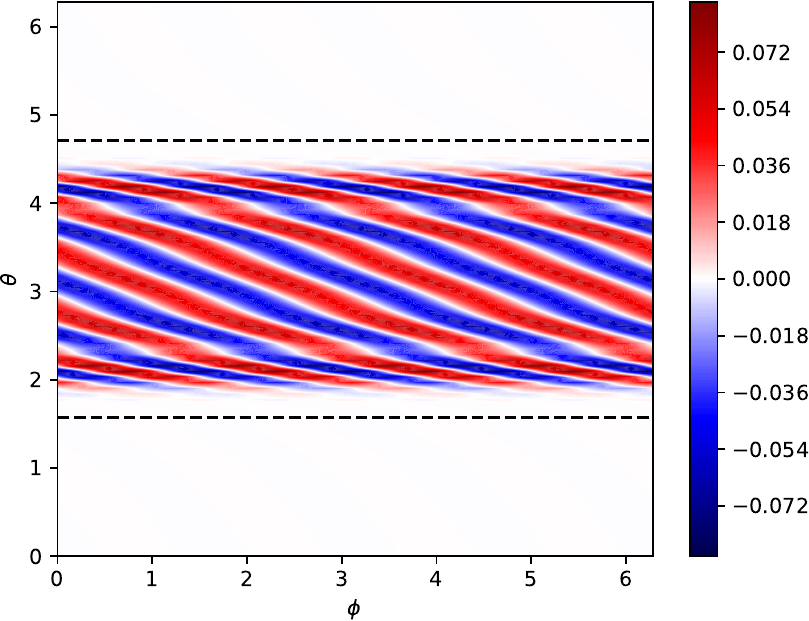} \\
    \includegraphics[width=0.32\linewidth]{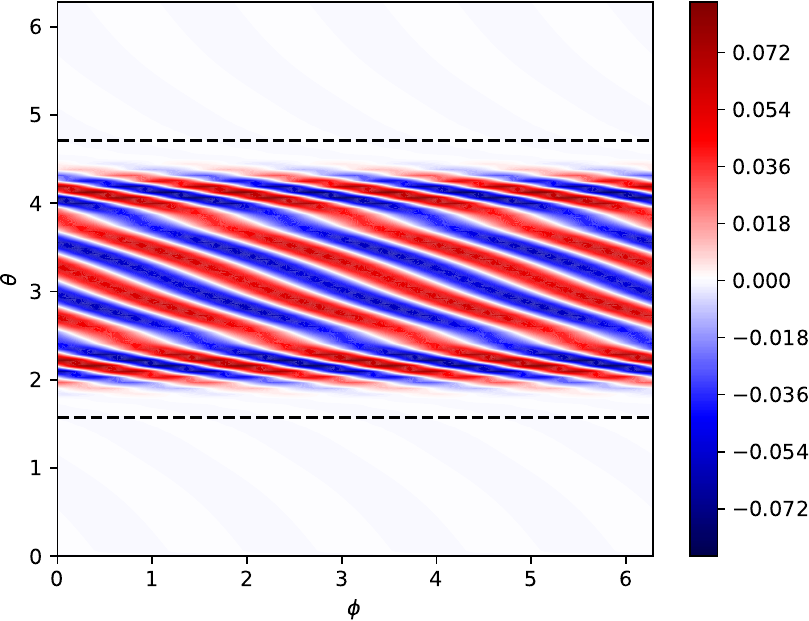} \includegraphics[width=0.32\linewidth]{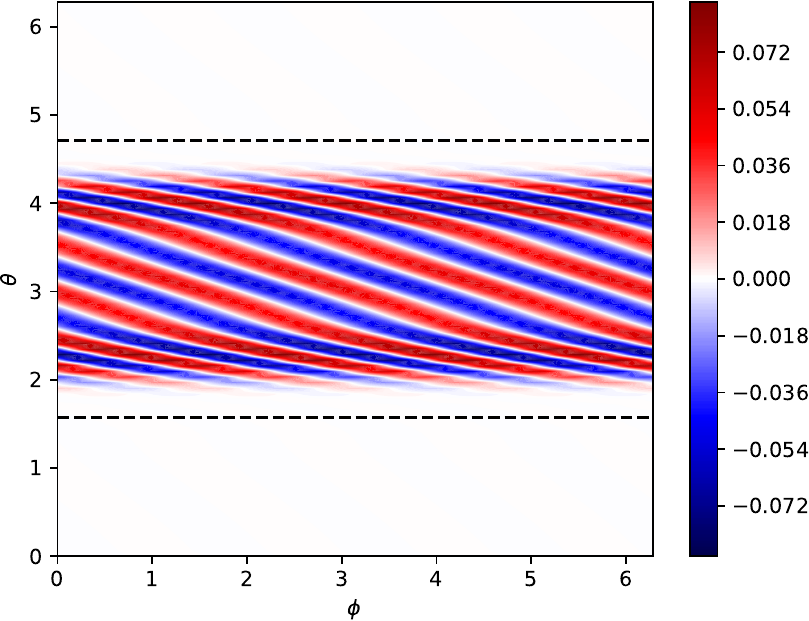} \includegraphics[width=0.32\linewidth]{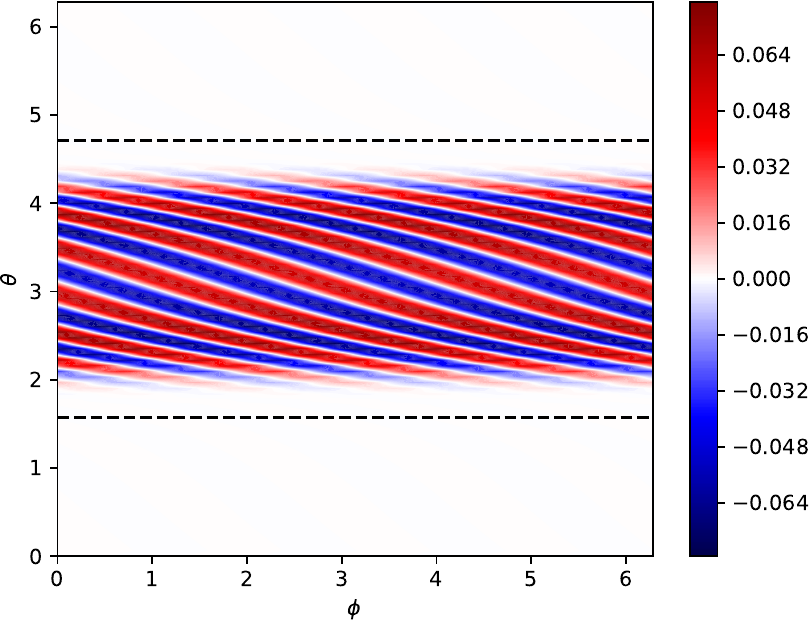}
    \caption{Several numerical solutions to equations \eqref{eq:2_bal_eqns} with $k_\alpha=4$ and 400 linear finite elements in the region $-3.25\pi R_0\leq l\leq 3.25\pi R_0$. The normalized residual $|\mathbf{Ax}|/\|\mathbf{A}\|_2$, where $\|\cdot\|_2$ is the spectral norm, is quite low, ranging from $\sim 2.0\times 10^{-3}$ (top left) to $\sim 3.8\times 10^{-3}$ (bottom right). The dashed lines correspond to the lines of zero Gaussian curvature at $\theta = \pi/2, 3\pi/2$, which is the boundary between the inboard and outboard sides.}
    \label{fig:nase_elm}
\end{figure}

\begin{figure}
    \centering
    \includegraphics[width=0.32\linewidth]{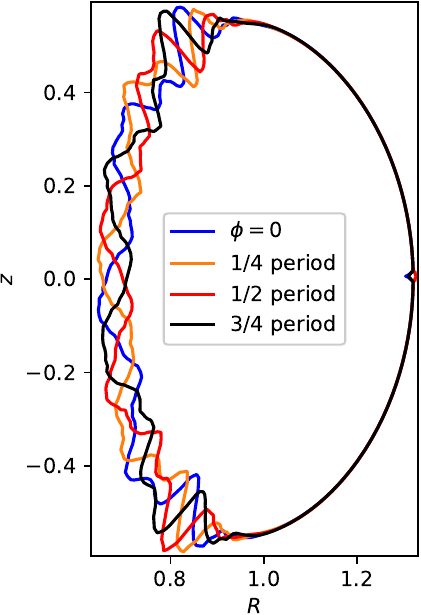} \includegraphics[width=0.32\linewidth]{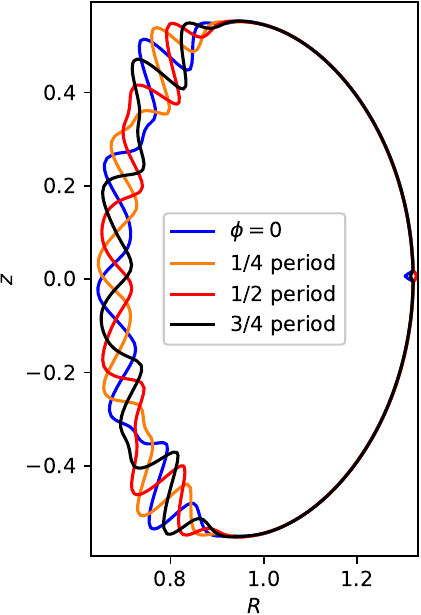} \includegraphics[width=0.32\linewidth]{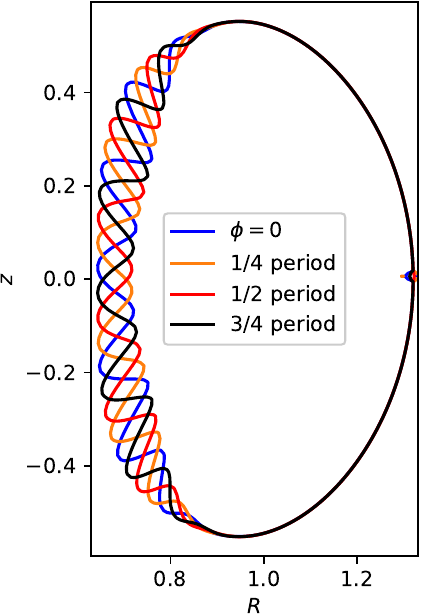} 
    \caption{Toroidal cross-sections for three different perturbations, corresponding to the solutions shown in the top left, top right and bottom right panels of Figure \ref{fig:nase_elm}, respectively. The cross sections are the contours of $\psi_0 + 0.1\psi_1$ corresponding to the outermost flux surface, where $\psi_1$ is normalized such that $|\mathbf{x}| = 1$, and $\mathbf{x}$ is the solution vector from equation \eqref{eq:eigvalform}.}
    \label{fig:nase_cross_sect}
\end{figure}

Next, we solve the same set of equations, but on an infinite ballooning-like domain, as discussed in section \ref{sec:bal_ridges}. Just as before, we fix the radial coordinate $\psi_0$ to be the outermost flux surface, however this time the coordinates $(\alpha_0,l)$ are used to parameterize the flux surface instead of $(\theta,\phi)$, where $\alpha_0$ is the field line label and $l$ is distance along the field line. We define $\alpha_0$ as the toroidal angle of the point where the given field line crosses the inboard midplane, and measure the distance $l$ from that point. Just like in section \ref{sec:bal_ridges}, and similarly to the example above, we keep only one Fourier mode in $\alpha_0$, with $k_\alpha = 4$. We then discretize the region of interest in $l$, $[-3.25\pi R_0,3.25\pi R_0]$, where $R_0 = 1$~m is the radius of the magnetic axis, using 400 1D linear finite elements. Mapped Zienkiewicz infinite elements in the Marques-Owen formulation are attached on both sides of the region of interest, enforcing the $\lim_{l\to\pm\infty} \{\Lambda_1,\psi_1\} = 0$ boundary condition. We refer the reader to chapter 4 of \citet{bettess1992infinite} for more details on Zienkiewicz infinite elements. Discretizing the equations using the Bubnov-Galerkin method, we again solve the eigenvalue problem \eqref{eq:eigvalform}; some numerically well-behaved (no subgrid oscillations and $\psi_1,\Lambda_1$ are small near the finite-infinite element boundary) solutions are shown in Figure \ref{fig:nase_elm}. We also show the cross-sections in Figure \ref{fig:nase_cross_sect}. Just as before, the solutions are localized and centered around $l=0$, the inboard midplane.

\subsection{Optimized near-axisymmetric configurations with QS in a volume}
\label{sec:Rogerio_numerics}

In paper I, we presented quasi-axisymmetric equilibria with various field periods ($n_{fp}=2,3..6$) obtained using the SIMSOPT code \cite{landreman2021simsopt} with axisymmetric tori as initial conditions. The details of the optimization procedure are provided in I. 

\begin{figure}
    \centering
    \includegraphics[width=0.315\linewidth]{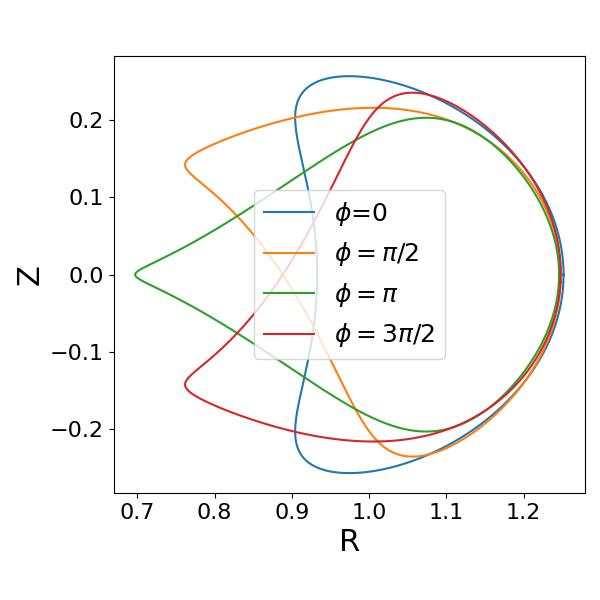}
    \includegraphics[width=0.33\linewidth]{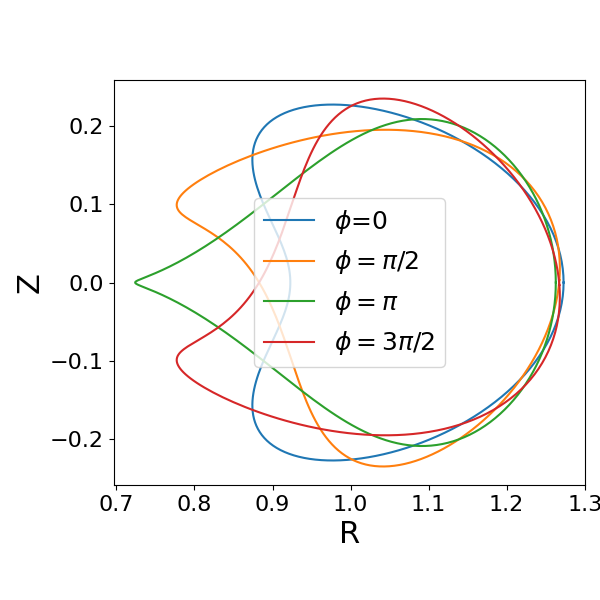}
    \includegraphics[width=0.315\linewidth]{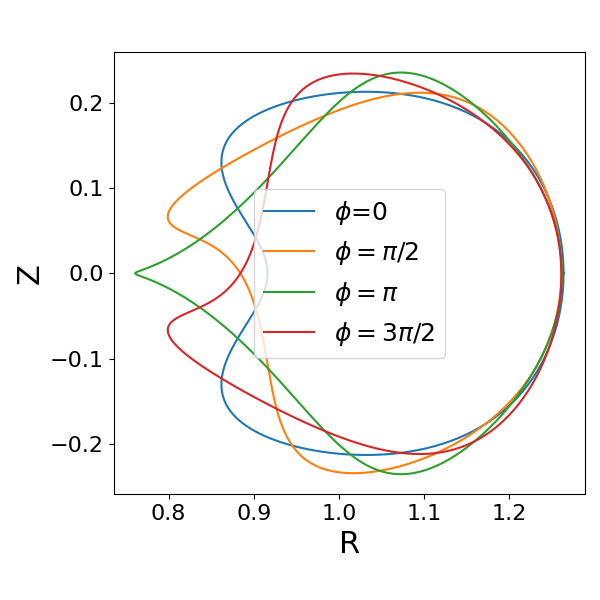}
    \caption{Poloidal cross-sections of the plasma boundary of quasi-axisymmetric configurations at toroidal angles $(2/n_{fp})\phi=0, \pi/2, \pi$ and $3\pi/2$ with $n_{fp}=3$ (left), $n_{fp}=4$ (middle) and $n_{fp}=6$ (right).}
    \label{fig:poloidal_nfp346}
\end{figure}

Figure \ref{fig:poloidal_nfp346} shows the poloidal cross sections of configurations with $n_{fp}=3,4$ and $6$ in cylindrical coordinates $(R,Z)$ at different cylindrical toroidal angles $\phi$.
Consistent with \cite{henneberg2024compact}, we find that while sharp ridges appear on the inboard side of the stellarator, the outboard side is almost axisymmetric with a nearly circular cross-section around the mid-board plane at poloidal angle $\theta=0$. Qualitatively the results or our linear perturbation results shown in Figure \ref{fig:poloidal_nfp346} agree with the fully nonlinear ridges shown in Figure \ref{fig:nase_cross_sect}. The main source of difference is that we performed a local analytical calculation on a single flux surface admitting only small radial perturbations from the ridges, whereas the numerically optimized cross sections allow for arbitrary ridge amplitudes.

\begin{figure}
    \centering
    \includegraphics[width=0.99\linewidth]{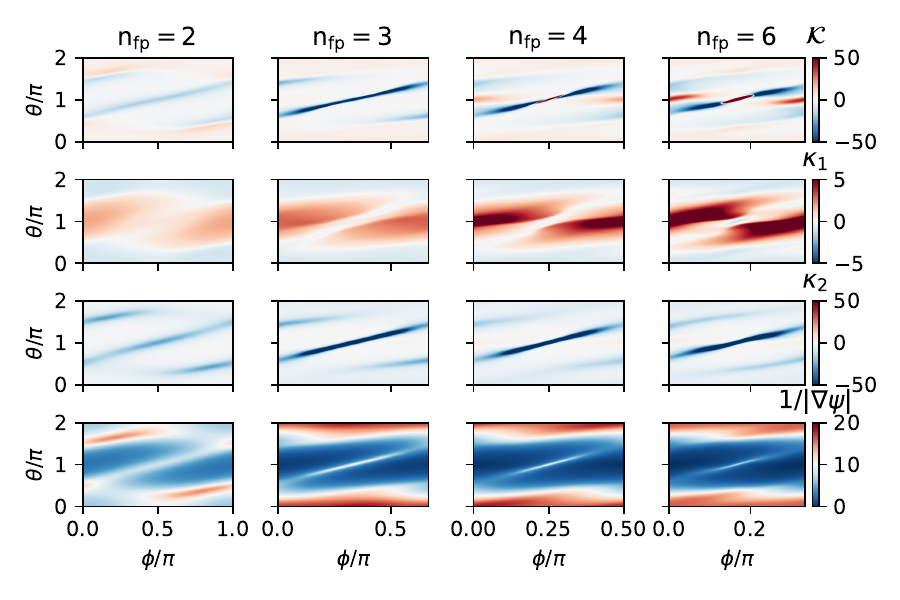}
    \caption{Gaussian curvature, $\cK$ (top), and principal curvatures $k_1$ (middle) and $k_2$ (bottom) on half a field period of the outer boundary of the optimized $n_{fp}=2, 3, 4$ and $6$ quasi-axisymmetric stellarators.}
    \label{fig:gaussian_curvatures_QA}
\end{figure}

Next, we explore the surface curvatures for the various field periods. In Figure \ref{fig:gaussian_curvatures_QA} we show the Gaussian curvature, $\cK$, and principal curvatures $k_1$ and $k_2$ on a field period of the outer boundary of the optimized $n_{fp}=2, 3, 4$ and $6$ quasi-axisymmetric stellarators.
In general, each one of these quantities increases its range of values with an increasing number of field periods.
We also show the values of $1/|\nabla \psi|$ at the boundary.

\begin{figure}
    \centering
    \includegraphics[width=0.8\linewidth]{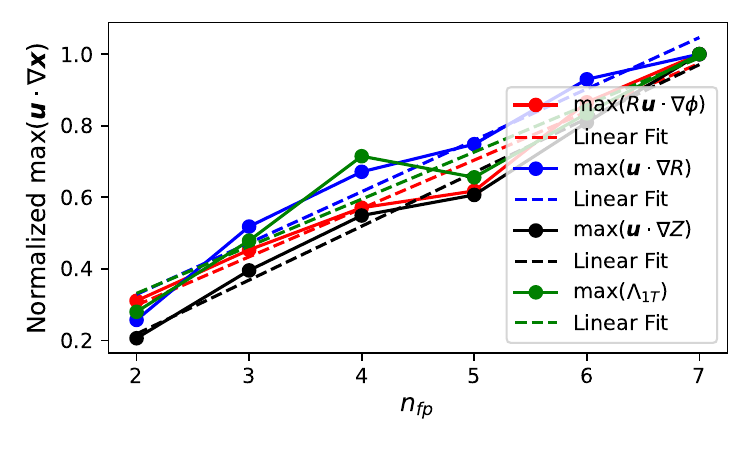}
    \caption{Maxima of the quasisymmetry vector $\mathbf u$ projected along the $\nabla \phi$ (red), $\nabla R$ (blue) and $\nabla Z$ (black) directions, and their respective linear fits (dashed), as a function of the number of field periods $n_{fp}$. The combination $\Lambda_1=R \mathbf u \cdot \nabla \psi_0/Z - Z \mathbf u \cdot \nabla \psi_0/R$ where $\psi_0$ is the axisymmetric counterpart of $\psi$ is also shown. Quantities are normalized to their value at $n_{fp}=7$.}
    \label{fig:u_dot_grad}
\end{figure}

In Figure \ref{fig:u_dot_grad} we show the maxima of the quasisymmetry vector $\mathbf u$ projected along the $\nabla \phi$ (red), $\nabla R$ (blue) and $\nabla Z$ (black) directions, and their respective linear fits (dashed), as a function of the number of field periods $n_{fp}$. The linear prediction of the $n_{fp}$ dependence is thus verified. 

\begin{figure}
    \centering
    \includegraphics[width=0.99\linewidth]{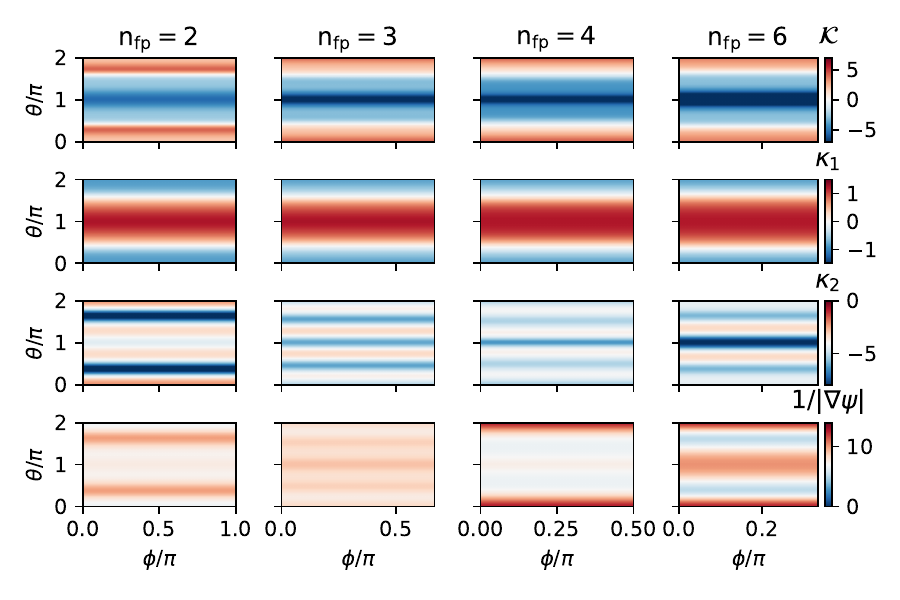}
    \caption{Gaussian curvature, $\cK$ (top), and principal curvatures $k_1$ (middle) and $k_2$ (bottom) on half a field period of the outer boundary of the axisymmetric counterparts, $\psi_0$ of the optimized $n_{fp}=2, 3, 4$ and $6$ quasi-axisymmetric stellarators.}
    \label{fig:gaussian_curvatures_QA_psi0}
\end{figure}

Next, we removed the non-axisymmetric components of the plasma boundary for each of the optimized solutions in this section to obtain $\psi_0$, i.e., we removed the toroidal mode number coefficients $n$ of the plasma boundary for each of the configurations optimized here.
The resulting curvatures are shown in Figure \ref{fig:gaussian_curvatures_QA_psi0}.

\begin{figure}
    \centering
    \includegraphics[width=\linewidth]{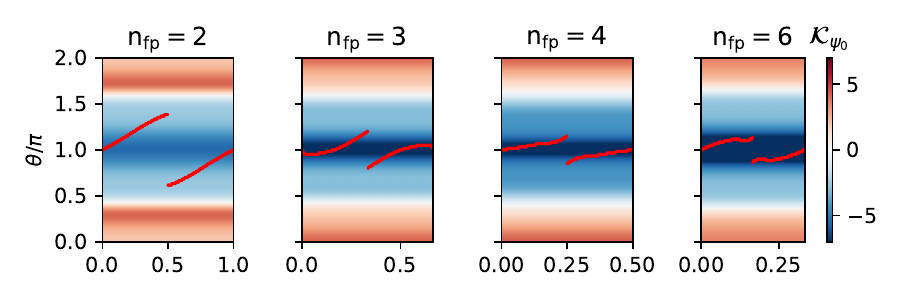}
    \caption{Location of the ridges, maxima of $k_2$ of the quasisymmetric configurations, superimposed with the Gaussian curvature of $\psi_0$. The ridges are observed to prefer areas with strongly negative Gaussian curvature, consistent with analytical predictions for vacuum configurations.}
    \label{fig:psi0K_ridges}
\end{figure}

The ridges, maxima of $k_2$ from the full solution, are superimposed with the Gaussian curvature of $\psi_0$ in Figure \ref{fig:psi0K_ridges}. As expected from the linear NASE theory, the ridges prefer the strongly negative Gaussian curvature of the axisymmetric background. 

\begin{figure}
    \centering
    \includegraphics[width=\linewidth]{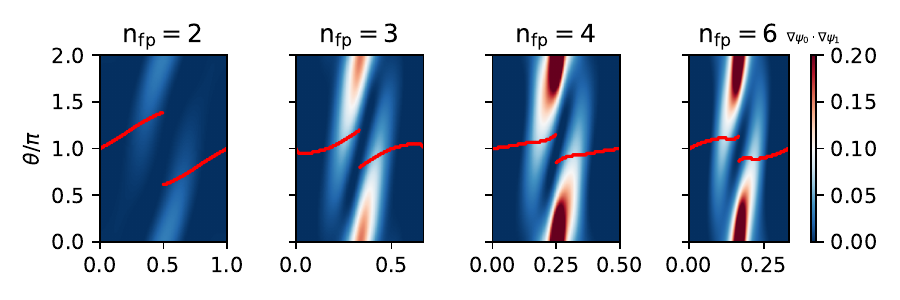}
    \caption{Location of the ridges, maxima of $k_2$ of the quasisymmetric configurations, superimposed on $\dl\psi_0\cdot \dl \psi_1$. We note that the ridge lines strongly avoid the large values of $\dl\psi_0\cdot \dl \psi_1$, consistent with the assumption of the ``ridge-conditions" \eqref{eq:ridge_conditions} in the analytical calculations. }
    \label{fig:psi0K_gradpsi0dotgradpsi1}
\end{figure}

Finally, in Figure \ref{fig:psi0K_gradpsi0dotgradpsi1}, we show that the important assumption that ridges are localized on the inboard side around regions where $\dl\psi_1\cdot\dl\psi_0$ is small holds in the fully nonlinear situation as well. In particular, we see that this assumption is violated in the outboard region where ridges are absent. Thus, the localization of the ridges as described by the analytical solutions of the ridge equations \eqref{eq:2_bal_eqns} obtained in Sections \ref{sec:Ridges_LargeN} and \ref{sec:bal_ridges} and numerical solutions obtained in \ref{sec:Nikita_numerics} is shown to be self-consistent.  

\section{Conclusions}
\label{sec:conclusions}

We have developed a systematic theory of ridges in quasiaxisymmetric magnetohydrostatic equilibrium with isotropic pressure close to axisymmetry. In particular, we focus on understanding the apparent localization of the sharp ridges to the inboard side as seen in \cite{plunk2018,plunk2020_near_axisymmetry_MHD,henneberg2024compact}. Our approach in this work is based on a perturbative expansion in the deviation from axisymmetry of the generalized quasisymmetric Grad Shafranov system of equations from \cite{burby2020}.

As is well-known, the system of quasisymmetric MHS is overdetermined in the isotropic pressure limit. Fortunately, the overdetermination is not a hindrance in the study of the sharp ridges, which are associated with regions of maximum flux expansion, or equivalently, locally coincide with the minimum of $|\dl\psi|$. Leveraging this key requirement, we can obtain a closed set of equations that describes the ridges. Further analytical progress is made with the help of a subsidiary expansion $N\gg 1$, where $N$ is the number of field. As a complementary approach to the large $N$ subsidiary, we utilize a ballooning-like eikonal formalism for ridges.

Our key results show that breaking axisymmetry while preserving quasiaxisymmetry leads to three-dimensional perturbations that are highly localized on the inboard side. We demonstrate the crucial role played by the Gaussian curvature of the flux surface in the localization of these perturbations. Ridges in perturbed axisymmetric tokamaks can only exist in regions where the Gaussian curvature of the axisymmetric background is non-positive. The localization effect of the ridges is very sensitive to the Gaussian curvature of the surface, but is not so sensitive to currents or pressure profile. The weak dependence on currents and pressure can be justified by the fact that near the ridges, $|\dl\psi|$ is small, so the equilibrium is closer to a force-free limit.

We verify our key analytical results by numerically solving the ridge equations. For the case of large $N$, we use a Fourier basis, whereas for the ballooning case, we use Zienkiewicz infinite elements. The background axisymmetric equilibrium is chosen to be an ITER-like Solov'ev model from \cite{pataki_cerfon_2013fastGS}. We also compare with fully three-dimensional compact vacuum QA configurations with sharp ridges, obtained using VMEC. In particular, we demonstrate the self-consistency and validity of the analytical assumptions using VMEC configurations.

The basic ideas of ridges aligning with a minimum of $|\dl\psi|$ and the preference for non-positive Gaussian curvature do not depend on closeness to axisymmetry. The perturbative approach was motivated primarily to make analytical progress on a complicated, nonlinear, overdetermined system of PDEs. The generalization to stellarators far from axisymmetry will be presented in a subsequent work. Finally, the physics-informed neural network (PINN) approach can be used to solve the overdetermined set of equations. Such models should help navigate the parameter space of compact flux-surface shapes with sharp ridges.


\section*{Acknowledgements}
The authors would like to thank Per Helander, Eduardo Rodriguez, Alan Kaptanoglu, Sophia Henneberg, Gabe Plunck, Josh Burby, Carl Sovinec, Jim Meiss, Joaquim Loizu, Chris Smiet, Bob Davies, Jos\'e-Luis Velasco, Felix Parra Diaz for stimulating discussions and helpful suggestions.

\section*{Funding}
This research was supported by a grant from the Simons Foundation/SFARI (560651, AB) and the Department of Energy Award No. DE-SC0024548 (until March 31, 2025). Some computations were performed on the HPC system Viper at the Max Planck Computing and Data Facility (MPCDF).

\section*{Declaration of interests}
The authors report no conflict of interest.

\section*{Data availability}
The data that support the findings of this study are available from the corresponding author upon reasonable request.

\appendix

\section{Equivalence of different forms of QS}
\label{app:eqv_QS}
The equivalence of different forms of QS has been discussed in several earlier works \citep{burby2020,constantin2021,rodrigGBC}. We compile the essential results in this Appendix to ensure the work remains self-contained.  

Starting from the weak form of QS, we observe that $\u$ can be expressed in terms of $\B$ using \eqref{eq:Bxu} as
\begin{align}
    \u =\frac{1}{B^2}((\u\cdot \B) \B -\B\times \dl \psi).
    \label{eq:u_in_terms_of_B}
\end{align}
From the condition \eqref{eq:uDB}, it follows that
\begin{align}
    \u \cdot \B =\frac{\B \times \dl \psi\cdot \dl B}{\BD B}.
    \label{eq:u_dot_B}
\end{align}
The 2-term for QS assuming ideal MHS is then equivalent to 
\begin{align}
    \u\cdot \B =F(\psi).
    \label{eq:udotB}
\end{align}
To prove \eqref{eq:udotB}, we now impose the $\dl\cdot \u=0$ condition from weak QS on the form of $\u$ in \eqref{eq:u_in_terms_of_B}. It follows that,
\begin{align}
    \dl \cdot \u = -\frac{\u \cdot \dl B^2}{B^2}+ \frac{1}{B^2}\lbr \dl\cdot ( (\u\cdot\B) \B)-\dl \cdot(B\times \dl \psi)\rbr
    =\frac{1}{B^2}\BD(\u\cdot\B),
\end{align}
from the QS condition $\uD B=0$ together with the ideal MHS conditions $\dl \cdot \B=0, \J=\dl\times \B, \J\cdot \dl \psi=0$. Thus, $\dl \cdot \u =0$ implies that $\BD(\u \cdot \B)=0$. Consequently, the 2-term form \eqref{eq:udotB} is satisfied since $(\u \cdot \B)$ must be a flux function $F(\psi)$ on irrational surfaces, and by continuity, on rational surfaces as well. Similarly, starting with the 2-term form and assuming the ideal MHS force balance, we can define
\begin{align}
    \u =\frac{1}{B^2}(F(\psi)\B -\B \times \dl \psi)
    \label{eq:u_given_B}
\end{align}
which satisfies $\uD B=0, \dl \cdot \u$ by construction. Taking the cross product with $\B$ we recover $\B\times \u =\dl \psi$ since $\BD \psi=0$. 

Next, we show the equivalence of the weak and the strong forms of QS. For a general force-balance condition, the strong form of QS is more restrictive than the weak form since \eqref{eq:strong_QS_Jxu} has more components compared to the single constraint $\dl\cdot \u=0$. However, for ideal MHS force balance, \eqref{eq:strong_QS_Jxu} has only one component in $\dl \psi$ direction. Hence, the weak and strong forms can be made equivalent by a suitable choice of a flux function described below. 

From the ideal MHS force balance \eqref{eq:ideal_MHS}, the current is of the form
\begin{align}
    \J=\jpl \B + \frac{\B\times \dl p}{B^2},
\end{align}
where the parallel component $\jpl$ is not yet determined. To determine $\jpl$, we impose $\dl \cdot \J=0$, which leads to the magnetic differential equation
\begin{align}
    \BD \jpl + \B \times \dl p\cdot \dl \lbr \frac{1}{B^2} \rbr=0.
    \label{eq:BDjpl}
\end{align}
Now, the QS condition $\uD B=0$, implies that 
\begin{align}
    \B \times \dl p\cdot \dl \lbr \frac{1}{B^2} \rbr=\BD\lbr F(\psi) \frac{p'(\psi)}{B^2} \rbr,
    \label{eq:QS_2term_1/B^2}
\end{align}
which allows us to solve for $\jpl$ by lifting $\BD$ from \eqref{eq:BDjpl}. Assuming irrational rotational transform such that $\BD$ is invertible and assuming continuity across rational surfaces, we get $\jpl$ up to a flux function $\mathfrak{J}(\psi)$
\begin{align}
    \jpl = \mathfrak{J}(\psi) -F(\psi) \frac{p'(\psi)}{B^2}.
    \label{eq:jpl_exp}
\end{align}
Using the form of $\u$ \eqref{eq:u_in_terms_of_B}, we can rewrite $\J$ as
\begin{align}
    \J= -p'(\psi)\u + \lbr \jpl +\frac{p' F}{B^2}\rbr \B = -p'(\psi)\u + \mathfrak{J}(\psi) \B
    \label{eq:J_u_B_reln}
\end{align}
Therefore, $\J$ satisfies both the ideal force-balance $\J\times \B=p'\dl \psi$, and the strong QS condition $\J \times \u +F'\dl \psi=0$, provided we make the choice
\begin{align}
    \mathfrak{J}(\psi)= -F'(\psi).
    \label{eq:Jbar_Fprime}
\end{align}
As shown in Appendices B and C of \cite{rodrigGBC}, weak QS has a gauge degree of freedom corresponding to the choice for the flux surface label. The $\J \times \u +F'\dl \psi=0$ condition is not gauge-invariant, so a choice of gauge is required. The gauge that leads to $\uD =\del_\phi$ is a natural choice, and it also leads to the choice \eqref{eq:Jbar_Fprime}.

\section{Derivation of the GGSE system \eqref{eq:Basic_QS_MHS_system} }
\label{app:Derivation_GGSE}
We  provide the necessary steps to derive the GGSE \eqref{eq:Basic_QS_MHS_system} starting with the ideal MHS equations and the weak QS conditions, which lead to
\begin{align}
    \B = \frac{1}{u^2}(F(\psi)\u +\u \times \dl \psi).
    \tag{\ref{eq:B_in_terms_of_u}}
\end{align}
First, we set the divergence of $\B$ to zero. Thus, $\dl\cdot (u^2 \B)= \BD u^2$. Now, $\dl\cdot (u^2 \B)$ from \eqref{eq:B_in_terms_of_u}, is given by
\begin{align}
    \dl \cdot (u^2 \B)&= \dl\cdot(F(\psi)\u) + \dl \cdot(\u\times \dl \psi) = \v \cdot \dl \psi,
\end{align}
upon using $\dl\cdot \u=0, \uD \psi=0$. Thus, we get 
\begin{align}
    \BD u^2 -\v\cdot \dl \psi=0.
    \tag{\ref{eq:Bdotw_is_0}}
\end{align}
From the definition $\w=\v\times \u +\dl u^2$, we find that \eqref{eq:Bdotw_is_0} is equivalent to $\w\cdot\B=0$. Thus, $\w$ must be orthogonal to $\B$.

Now, we take the curl of $\B$ to compute $\J$ from \eqref{eq:B_in_terms_of_u} using
\begin{align}
    \J&= \dl \times \lbr \frac{1}{u^2}(u^2 \B)\rbr = \frac{1}{u^2}\lbr \B \times \dl u^2 + \dl \times (u^2 \B) \rbr \nonumber\\
    &=\frac{1}{u^2}\lbr \B \times \dl u^2 +F\v -F' \u\times \dl \psi+\dl \times (\u\times \dl \psi)\rbr \label{eq:J_terms}.
\end{align}
Now, we evaluate each one of these terms. Firstly, from 
$\w=\v\times \u +\dl u^2$ and $\u\cdot \B=F$, it follows that
\begin{align}
\B \times \w = \B \times \dl u^2+F \v -(\v \cdot \B) \u,\quad \v\cdot \B = \frac{1}{u^2}(F (\u\cdot \v) -(\u \times \v) \cdot \dl \psi).
    \label{eq:Bxw_form}
\end{align}
Thus,
\begin{align}
    \B\times \dl u^2 + F \v = \B \times \w + \frac{1}{u^2}(F (\u\cdot \v) -(\u \times \v) \cdot \dl \psi)\u.
    \label{eq:Bxdlu2_plus_Fv}
\end{align}
Next, the term $-(F'/u^2) \u\times \dl \psi$ will be written in terms of $\B$ as
\begin{align}
    -\frac{F'}{u^2} \u\times \dl \psi = -F' \B + \frac{1}{u^2}( F F') \u.
    \label{eq:Fpu2}
\end{align}
Finally, we evaluate the curl of $\u \times \dl \psi$ to find
\begin{align}
    \dl \times (\u \times \dl \psi )= (\nabla^2  \psi) \u + \dl \psi \cdot \dl \u - \uD \dl \psi
    \label{eq:curl_uxdlpsi}
\end{align}
Substituting \eqref{eq:Bxdlu2_plus_Fv}, \eqref{eq:Fpu2} and \eqref{eq:curl_uxdlpsi} in \eqref{eq:J_terms} and collecting various terms we find
\begin{align}
    \J &= -F'\B -p' \u + \frac{1}{u^2}\J_w + \frac{1}{u^2}J_{\text{GGS}}\; \u ,\label{eq:J_form_full}\\
    \J_w &= \B\times \w - \lbr \uD \dl \psi -\dl\psi\cdot \dl \u\rbr, \nonumber\\
    J_{\text{GGS}} &= \nabla^2  \psi -\frac{\u \times \v}{u^2}\cdot \dl \psi +\frac{\u\cdot \v}{u^2}F + FF' +u^2 p'. \nonumber
\end{align}

We next discuss the structure of the $\J_w$ vector by taking components in $(\u,\dl \psi, \B)$ directions. The following vector identities will prove to be useful
\begin{subequations}
    \begin{align}
    (\bm{A}\times \bm{C})\cdot \dl \times \B &= \bm{A}\cdot \dl \B \cdot \bm{C}- \bm{C}\cdot \dl \B \cdot \bm{A} \label{eq:vec_id1},\\
    \dl(\bm{A}\cdot \B)&= \dl \bm{A}\cdot \B + \dl \B\cdot \bm{A} \nonumber\\
    &= \bm{A} \times (\dl \times \B)+\B \times (\dl \times \bm{A}) +\bm{A}\cdot \dl \B + \BD \bm{A}
    \label{eq:vec_id2},\\
    \dl \B \cdot \bm{A}&=\bm{A}\cdot \dl \B + \bm{A}\times (\dl \times \B) \label{eq:vec_id3}.
\end{align}
\label{eq:vec_ids}
\end{subequations}
Dotting $\J_w$ with $\u$ and using $\B\times \u =\dl \psi$, together with the identities \eqref{eq:vec_ids}, we get
\begin{align}
    \J_w \cdot \u &= -\w \cdot \dl \psi +\dl \psi \cdot \dl u^2/2 -\u\cdot \dl\dl \psi\cdot \u, \nonumber\\
    &= -\w \cdot \dl \psi + \lbr \dl u^2/2 +\uD \u\rbr \cdot \dl \psi \label{eq:Ju.u}\\
    &= -\w \cdot \dl \psi +(\dl u^2 + \v\times \u )\cdot \dl \psi=0. \nonumber
\end{align}
Similarly, 
\begin{align}
    \J_w \cdot \B &= \dl\psi \cdot \dl \u \cdot \B -\uD  \dl \psi\cdot \B \nonumber\\
    &=\dl \psi \cdot \dl \u \cdot \B-\BD \dl \psi\cdot \u=\dl \psi \cdot \dl \u \cdot \B +\B \cdot \dl \u \cdot \dl \psi  \label{Jw.B}\\
    &= 2\dl\psi\cdot \tnsr{S}\cdot \B. \nonumber
\end{align}
Finally, using $$(\B\times \w) \cdot \dl\psi=(\B\times \w) \cdot (\B\times \u)=B^2 (\u\cdot \w)=B^2 \uD u^2=\uD(uB)^2-u^2 \uD B^2$$ since $\w\cdot \B=0$, we find
\begin{align}
    \J_w \cdot \dl \psi &= (\B \times \w)\cdot \dl \psi- \uD|\dl \psi|^2/2 +\dl \psi \cdot \dl \u \cdot \dl \psi \nonumber\\
    &= -u^2\uD B^2+\uD(|\dl\psi|^2+F^2) - \uD|\dl \psi|^2/2 -\dl \psi \cdot \dl \dl \psi \cdot \u  \nonumber\\
    & =-u^2\uD B^2 + \uD \dl\psi\cdot \dl \psi -\dl \psi\cdot \dl \dl \psi\cdot \u
    \label{Jw.dl_psi}\\
    &= -u^2 \uD B^2 +(\u \times \dl \psi).(\dl \times \dl \psi) = -u^2 \uD B^2. \nonumber
\end{align}
Therefore, 
\begin{align}
    \J_w &=\B \frac{1}{B^2}(\J_w \cdot \B) +  \dl \psi\frac{1}{|\dl \psi|^2}(\J_w \cdot \dl\psi)\\
    \J_w \cdot \B &=2\dl\psi\cdot \tnsr{S}\cdot \B,\nonumber\\
    \J_w \cdot \dl \psi &= -u^2 \uD B^2. \nonumber
\end{align}

\section{Derivation of the first-order NASE system }
\label{app:Derivation_NASE}
The derivation of the first-order NASE system follows from a straightforward linearization of the QS-MHD system \eqref{eq:Basic_QS_MHS_system} about an axisymmetric equilibrium. Therefore, we only give the essential steps. 

The first-order system is given by
\begin{subequations}
    \begin{align}
    &\u_0\cdot \dl \psi_1 + \u_1\cdot \dl\psi_0=0,\\
    &\u_0\cdot \dl (2 \B_0\cdot \B_1)+ \u_1 \cdot \dl B_0^2 =0, \\
    &\dl \cdot \u_1 =0, \\
    &-\B_0\cdot \dl (2 \u_0\cdot \u_1)- \B_1 \cdot \dl R^2 + \v_0\cdot \dl \psi_1 +\v_1 \cdot \dl \psi_0=0,\\
    &\dl \psi_0 \cdot \dl \u_1 \cdot \B_0 + \B_0\cdot \dl \u_1 \cdot \dl \psi_0 =0,\\
    &R^2 \dl \cdot \lbr \frac{1}{R^2}\dl \psi_1 \rbr  +\psi_1 \lbr (F')^2 +F F'' + R^2 p''\rbr + 2 \u_0\cdot \u_1 p' \\
   &+(\u_0\cdot \v_1 +\u_1\cdot \v_0)\frac{F}{R^2}-\frac{1}{R^2}\lbr \u_0\times \v_1 +\u_1\times\v_0 -2 \frac{\u_0\cdot \u_1}{R}2\Rh\rbr \cdot \dl\psi_0 =0, \nonumber
\end{align}
 \label{eq:O(ep)}
\end{subequations}
where,
\begin{align}
    \B_1 &= -2 \frac{\u_0 \cdot \u_1}{R^2}\B_0 +\frac{1}{R^2}\lbr F \u_1 +\psi_1 F' \u_0 + \u_0\times \dl \psi_1 +\u_1 \times \dl \psi_0\rbr,\nonumber\\
    \v_1 &= \dl \times \u_1, \quad 
    \B_0\cdot \dl = \frac{F}{R^2}\del_\phi +\frac{1}{R}\{\psi_0,\;\; \}_{(z,R)}. \label{eq:B1_v1}
\end{align}

Also, note that $\B_1$ follows directly from the expansion of \eqref{eq:B_form} to first-order. 

The divergence of $\u_1$ expressed in cylindrical coordinates gives \eqref{eq:div_u_cyl}. To get \eqref{eq:uDB_cyl} and \eqref{eq:div_B_cyl}, we use the following expressions that follow from the expression for $\B_1,\v_1$ given in \eqref{eq:B1_v1} and \eqref{eq:u1_v1_components},
\begin{subequations}
\begin{align}
    \B_0\cdot \B_1 &= \frac{1}{R^2}\lbr -\u_0\cdot \u_1 B_0^2 +\psi_1 F F' +\dl \psi_0\cdot \dl \psi_1 \rbr\\
    \B_1 \cdot \dl R^2 &= \frac{2 F}{R}u_{1R}-\frac{2}{R}\u_{1\phi} \del_z\psi_0 +2\del_z \psi_1 \\
    \v_1\cdot \dl \psi_0 &= \frac{1}{R}\{\psi_0, R u_{1\phi}\} +\frac{1}{R}\del_\phi \lbr u_{1z}\del_R \psi_0 -u_{1R}\del_z \psi_0 \rbr.
\end{align}
    \label{eq:B1_v1_related}
\end{subequations}

The expression for $\J_w\cdot \B=0$ to first-order, i.e., $\dl\psi_0\cdot \tnsr{S}_1\cdot \B_0=0$ is given by \eqref{eq:JwB_cyl}, which follows from
\begin{align}
\dl\psi_0\cdot\tnsr{S}_1\cdot \B_0 &=\dl\psi_0 \cdot \dl \u_1 \cdot \B_0 +\B_0 \cdot \dl \u_1 \cdot \dl\psi_0 \nonumber\\
    &= \dl\psi_0 \cdot \dl \lbr \u_1 \cdot \B_0 \rbr +\B_0 \cdot \dl (\u_1 \cdot \dl \psi_0)-\dl \psi_0 \cdot \dl \B_0 \cdot \u_1 -\B_0\cdot \dl \dl\psi_0 \cdot \u_1 \nonumber\\
    &=\dl\psi_0 \cdot \dl \lbr \u_1 \cdot \B_0 \rbr -\B_0 \cdot \dl (\u_0 \cdot \dl \psi_1)-\dl \psi_0 \cdot \dl \B_0 \cdot \u_1 -\u_1\cdot \dl \dl\psi_0 \cdot \B_0 \nonumber\\
    &=\dl\psi_0 \cdot \dl \lbr \u_1 \cdot \B_0 \rbr -\B_0 \cdot \dl (\u_0 \cdot \dl \psi_1)-\dl \psi_0 \cdot \dl \B_0 \cdot \u_1 +\u_1\cdot \dl \B_0  \cdot  \dl\psi_0 \nonumber\\
    &=\dl\psi_0\cdot \dl\lbr \u_1 \cdot \B_0\rbr-\B_0\cdot \dl (\u_0\cdot \dl \psi_1) -\u_1 \cdot \J_0 \times \dl \psi_0 \label{eq:psiSB_form}
\end{align}

Similarly, the alternative expression for $\dl\cdot \B_1 =0$ given by $\u_0\cdot \tnsr{S}_1\cdot \B_0=0$ takes the form
\begin{align}
    \u_0\cdot\tnsr{S}_1\cdot \B_0 &=\u_0 \cdot \dl \u_1 \cdot \B_0 +\B_0 \cdot \dl \u_1 \cdot \u_0 \nonumber\\
    &=\u_0 \cdot \dl \lbr \u_1 \cdot \B_0\rbr +\B_0\cdot \dl \lbr \u_0\cdot \u_1 \rbr -(\B_0 \cdot \dl \u_0+\u_0\cdot\dl\B_0) \cdot \u_1\nonumber\\
    &=\u_0 \cdot \dl \lbr \u_1 \cdot \B_0\rbr +\B_0\cdot \dl \lbr \u_0\cdot \u_1 \rbr -(\J_0\times \u_0 +\dl(\u_0\cdot \B_0)-(\v_0\times \B_0)) \cdot \u_1\nonumber\\
    &=\u_0 \cdot \dl \lbr \u_1 \cdot \B_0\rbr +\B_0\cdot \dl \lbr \u_0\cdot \u_1 \rbr -\u_1 \cdot \lbr \v_0\times \B_0 \rbr \nonumber\\
    &=\del_\phi \lbr \u_1 \cdot \B_0\rbr +\B_0\cdot \dl (R\u_{1\phi})+\frac{2F}{R}u_{1R}-\frac{2}{R}\del_z \psi_0 u_{1\phi}\label{eq:uSB_alt}
\end{align}
We have repeatedly used the vector identities \eqref{eq:vec_ids} and $\J_0=-p' \u_0-F' \B_0$. In terms of $\Lambda_{1i}$'s, \eqref{eq:psiSB_form} and \eqref{eq:uSB_alt} yields \eqref{eq:psiSB_uSB_alt_L_form}.

The GGSE \eqref{eq:GGS_cyl} follows from simplifying the GGSE from \eqref{eq:O(ep)} using

    \begin{align}
        &\nabla^2 \psi_1 -\frac{\u_0\times \v_0}{R^2}\cdot \dl\psi_1 = \nabla^2 \psi_1 - \frac{2}{R}\dl R \cdot \dl \psi_1 = R^2\dl \cdot \lbr \frac{1}{R^2}\dl \psi_1\rbr, \nonumber \\
        &  R^2\dl \cdot \lbr \frac{1}{R^2}\dl \psi\rbr=\dlts \psi_1 +\frac{1}{R^2}\del_\phi^2 \psi_1 ,\nonumber\\
        & -\frac{1}{R^2}\lbr \u_0\times \v_1 +\u_1\times \v_0 -2 \u_0\cdot \u_1\frac{\u_0\times \v_0}{R^2} \rbr \cdot \dl \psi_0 =\v_1\cdot \dl\phi \times \dl \psi_0 +\frac{2u_{1\phi}}{R^2}\del_R\psi_0, \nonumber\\
        &\frac{F}{R^2}\lbr \u_1 \cdot \v_0 + \u_0 \cdot \v_1\rbr = F \v_1 \cdot \dl \phi +\frac{2F}{R^2}u_{1z},\nonumber\\
       &\v_1\cdot \B_0 = \v_1\cdot \lbr F\dl\phi +\dl \phi \times \dl \psi_0\rbr .  \label{eq:GGSE_ids}
    \end{align}

Further simplification of the GGSE can be carried out by noting that
\begin{align}
    \v_1\cdot \B_0&= \frac{F}{R} \lbr \del_z u_{1R}-\del_R u_{1z}\rbr+\frac{1}{R^2}\del_z\psi_0\lbr \del_\phi u_{1z}-\del_z (R u_{1\phi})\rbr -\frac{1}{R^2}\del_R\psi_0\lbr -\del_\phi u_{1R}+\del_R (R u_{1\phi})\rbr \nonumber\\
    &= \frac{F}{R} \lbr \del_z u_{1R}-\del_R u_{1z}\rbr-\frac{1}{R^2}\dl\psi_0\cdot \dl (R u_{1\phi})+\frac{1}{R^2}\del_\phi \lbr u_{1z}\del_z \psi_0 +u_{1R}\del_R \psi_0\rbr\nonumber\\
    &=\del_\phi\left[\frac{F}{R}\lbr \del_z \Lambda_{1R}-\del_R \Lambda_{1z}\rbr -\frac{1}{R^2}\dl\psi_0\cdot \dl (R\Lambda_{1\phi})\right] -\frac{1}{R^2}\del_\phi^2\psi_1,
\end{align}
where in the final step we have used the equation $\uD\psi=0$ to first-order \eqref{eq:uDpsi_cyl}.
Using 
Therefore, the GGSE can be written as
\begin{align}
    \dlts \psi_1 &+\lbr {F'}^2 + F F'' + R^2 p''  \rbr \psi_1  \nonumber\\
    &+\del_\phi\lbr \frac{2\Lambda_{1\phi}}{R}\lbr R^2 p'+\frac{1}{R}\del_R\psi_0\rbr +2 \frac{F}{R^2}\Lambda_{1z}\rbr
    + v_{1B}=0,
    \label{eq:Dlts_psi_1_form_of_GGSE}
\end{align}
where we have defined
\begin{align}
    v_{1B}  \equiv  \del_\phi\left[\frac{F}{R}\lbr \del_z \Lambda_{1R}-\del_R \Lambda_{1z}\rbr -\frac{1}{R^2}\dl\psi_0\cdot \dl (R\Lambda_{1\phi})\right], \quad 
\end{align}

In the cylindrical coordinates coordinates, the first-order NASE system thus takes the form
\begin{subequations}
\begin{align}
&\del_\phi \psi_1 +\lbr u_{1R}\del_R +u_{1z}\del_z \rbr \psi_0=0,\label{eq:uDpsi_cyl}\\
&\lbr u_{1R}\del_R +u_{1z}\del_z \rbr B_0^2 + \frac{2}{R^2}\del_\phi \lbr -B_0^2 R u_{1\phi} + \psi_1 FF' +\dl \psi_1 \cdot \dl \psi_0 \rbr=0,
\label{eq:uDB_cyl}\\
&\del_\phi u_{1\phi}+\del_R (R u_{1R})+\del_z (R u_{1z})=0,\label{eq:div_u_cyl}\\
&\B_0\cdot \dl (2u_{1\phi})- \frac{1}{R}\lbr \v_1 \cdot \dl \psi_0 -\frac{2 F}{R} u_{1R}\rbr =0,\label{eq:div_B_cyl}\\
&\dl\psi_0 \cdot \dl\lbr \u_1\cdot \B_0\rbr -\B_0\cdot \dl(\del_\phi \psi_1)-\u_1\cdot \lbr \J_0\times \dl \psi_0\rbr=0,\label{eq:JwB_cyl}\\
&R^2 \dl \cdot \lbr \frac{1}{R^2}\dl \psi_1 \rbr +\psi_1 \lbr {F'}^2 +F F'' + R^2 p''\rbr +\v_1 \cdot \B_0\nonumber\\
&+\frac{2}{R^2}\lbr u_{1\phi}\lbr R^3 p' +\del_R \psi_0\rbr +F u_{1z} \rbr=0,\label{eq:GGS_cyl}
\end{align}
\label{eq:O(ep))_cyl}
\end{subequations}
 with
 \begin{align}
 \v_1 \cdot \dl \psi_0 =&\frac{1}{R}\del_R\psi_0\lbr \del_\phi u_{1z}-\del_z (R u_{1\phi})\rbr +\frac{1}{R}\del_z\psi_0\lbr -\del_\phi u_{1R}+\del_R (R u_{1\phi})\rbr, \nonumber \\
 \v_1 \cdot \B_0 =&\frac{1}{R^2}\del_z\psi_0\lbr \del_\phi u_{1z}-\del_z (R u_{1\phi})\rbr -\frac{1}{R^2}\del_R\psi_0\lbr -\del_\phi u_{1R}+\del_R (R u_{1\phi})\rbr\nonumber\\
 &+\frac{F}{R} \lbr \del_z u_{1R}-\del_R u_{1z}\rbr,\label{eq:v1_u1_J0_ids}\\
 \u_1\cdot \B_0 =& \frac{F}{R}u_{1\phi}+\frac{1}{R}\lbr \del_z \psi_0\; u_{1R}-\del_R \psi_0\; u_{1z} \rbr \nonumber \\
 \J_0\times \dl \psi_0 =& \frac{1}{R^2}\lbr \dlts \psi_0 \u_0 \times \dl \psi_0 +|\dl \psi_0|^2 F' \u_0 \rbr. \nonumber
 \end{align}

In terms of $\Lambda_{1i}$'s we have
\begin{align}
\v_1\cdot\dl\psi_0=&\frac{1}{R}\del_\phi \lbr -\del_\phi \Lambda_1+\{\psi_0,R\Lambda_{1\phi}\}_{(z,R)} \rbr,\nonumber\\
\u_1\cdot\B_0 =&\del_\phi\lbr \frac{1}{R}(F \Lambda_{1\phi}+\Lambda_1) \rbr,\quad
 \v_1 \cdot \B_0 =-\frac{1}{R^2}\del_\phi^2\psi_1 +v_{1B},\nonumber\\
 v_{1B}  =&  \del_\phi\left[\frac{F}{R}\lbr \del_z \Lambda_{1R}-\del_R \Lambda_{1z}\rbr -\frac{1}{R^2}\dl\psi_0\cdot \dl (R\Lambda_{1\phi})\right].\label{eq:v1dotpsi_and_B0_L}
 \end{align}

With the definitions \eqref{eq:Lambda_def}, \eqref{eq:v1dotpsi_and_B0_L}, and  the expressions \eqref{eq:v1_u1_J0_ids}, the first-order NASE system \eqref{eq:O(ep))_cyl} reads
\begin{subequations}
\begin{align}
&\psi_1 +\lbr \Lambda_{1R}\del_R +\Lambda_{1z}\del_z \rbr \psi_0=0,\label{eq:uDpsi_cyl_L}\\
&\lbr \Lambda_{1R}\del_R +\Lambda_{1z}\del_z \rbr B_0^2 + \frac{2}{R^2} \lbr -B_0^2 R \del_\phi \Lambda_{1\phi} + \psi_1 FF' +\dl \psi_1 \cdot \dl \psi_0 \rbr=0,
\label{eq:uDB_cyl_L}\\
&\del_\phi \Lambda_{1\phi}+\del_R (R \Lambda_{1R})+\del_z (R \Lambda_{1z})=0,\label{eq:div_u_cyl_L}\\
&\B_0\cdot \dl (2 \Lambda_{1\phi})-\frac{1}{R^2}\{\psi_0,R\Lambda_{1\phi}\}+\frac{2 F}{R^2} \Lambda_{1R} +\frac{1}{R^2}\del_\phi \Lambda_1  =0,\label{eq:div_B_cyl_L}\\
& \B_0\cdot \dl \psi_1 -\dl\psi_0 \cdot \dl \lbr \frac{1}{R}(F \Lambda_{1\phi}+\Lambda_1)\rbr +|\dl\psi_0|^2 \frac{F'}{R}\Lambda_{1\phi} + \frac{\dlts \psi_0}{R} \Lambda_1=0 \label{eq:JwB_cyl_L}\\
&\dlts \psi_1 +\psi_1 \lbr {F'}^2 +F F'' + R^2 p''\rbr +v_{1B} +\frac{2}{R^2}\del_\phi\lbr R\Lambda_{1\phi}\lbr R^2 p' +\frac{1}{R}\del_R \psi_0\rbr +F \Lambda_{1z} \rbr=0
\label{eq:GGS_cyl_L}
\end{align}
\label{eq:O(ep))_cyl_L}
\end{subequations}

Note that we have lifted the $\del_\phi$ operator in multiple places and have not added any ``integration constants" of the axisymmetric form $f_1(R,z)$. This can be done because we assume all the axisymmetric terms come from the background and the first-order perturbations are all essentially nonsymmetric and periodic in $\phi$.

\section{Analysis of the structure of the first-order NASE system \eqref{eq:O(ep))_cyl_L}}
\label{app:structure_preserving}
We now formulate an approach to understand the overall algebraic structure of the equations and to obtain consistency conditions. 

It is convenient to group the system of equations \eqref{eq:O(ep))_cyl_L} as
\begin{enumerate}[I. ]
    \item $1.\;\;\Lambda_{1R}\del_R\psi_0 + \Lambda_{1z}\del_z\psi_0=-\psi_1,\;\; 2.\;\; \Lambda_{1R}\del_z\psi_0 - \Lambda_{1z}\del_R\psi_0=\Lambda_1$
    \item $1.\;\;\dl \cdot \u=0, \qquad \qquad \qquad\qquad \;\; 2.\;\;\uD B=0 \;$ ( $\Leftrightarrow \J_w \cdot \dl\psi=0$ )
    \item $1.\;\;\dl\cdot \B =0\; (\Leftrightarrow \u\cdot\tnsr{S}\cdot \B=0), \;\; 2.\;\;\J_w \cdot \B =0$ (  $\Leftrightarrow \dl\psi \cdot \tnsr{S}\cdot \B=0$ )
    \item $J_{\text{GGS}}=0$
\end{enumerate}
Group I consists of $\uD\psi=0$ to first order given by \eqref{eq:uDpsi_cyl_L}, together with the definition of $\Lambda_1$ from \eqref{eq:Lambda_def}. Group II consists of $\dl\cdot u=0$ and $\uD B=0$ to first-order given by \eqref{eq:div_u_cyl_L} and \eqref{eq:uDB_cyl_L}. Similarly, Group III is formed from the first-order equations \eqref{eq:div_B_cyl_L} and \eqref{eq:JwB_cyl_L}. Finally, Group IV is the first-order GGSE \eqref{eq:GGS_cyl_L}.

We now discuss each of these groups systematically in the following, and then summarize our analysis.

\subsection{Group I}
From Group I equations, we can solve for $(\Lambda_{1R},\Lambda_{1z})$ in terms of $\psi_1$ and $\Lambda_1$ as 
\begin{align}
    \begin{pmatrix}
    \Lambda_{1R}\\\Lambda_{1z}
    \end{pmatrix}
    =\tnsr{M_I}
    \begin{pmatrix}
    -\psi_{1}\\ +\Lambda_1
    \end{pmatrix},\qquad \tnsr{M_I}\equiv \frac{1}{|\dl \psi_0|^2}
    \begin{pmatrix}
    \del_R \psi_0 \quad +\del_z \psi_0 \\ \del_z \psi_0 \quad -\del_R \psi_0 
    \end{pmatrix}.
    \label{eq:I_sol}
\end{align}
The following identities obtained from \eqref{eq:I_sol} will be useful later on.
\begin{subequations}
    \begin{align}
   &\lbr \Lambda_{1R}\del_R + \Lambda_{1z}\del_z \rbr = \frac{1}{|\dl\psi_0|^2}\lbr-\psi_1\dl\psi_0\cdot \dl +\Lambda_1\{\psi_0,\;\; \}\rbr,
   \label{eq:Lambda_dot_grad}\\
   &\lbr \Lambda_{1R}\del_R + \Lambda_{1z}\del_z \rbr(R B_0)^2= -\psi_1 \lbr 2 FF' + \frac{\dl\psi_0\cdot\dl|\dl\psi_0|^2}{|\dl\psi_0|^2}\rbr + \frac{\Lambda_1}{|\dl\psi_0|^2}\{\psi_0,|\dl\psi_0|^2\}\\
  &\del_R (R\Lambda_{1R}) + \del_z(R\Lambda_{1z}) = \frac{R}{|\dl \psi_0|^2}\lbr -\dl \psi_0\cdot \dl \psi_1 +\{\psi_0,\Lambda_1\}\rbr-\cA_1\psi_1 -\cA_2 \Lambda_1,\\
  &\cA_1 \equiv\frac{R}{|\dl \psi_0|^2}\lbr \nabla^2 \psi_0 -\frac{\dl\psi_0\cdot \dl |\dl\psi_0|^2}{ |\dl\psi_0|^2}\rbr, \;\; \cA_2 \equiv -\left\{\psi_0,\frac{R}{|\dl \psi_0|^2}\right\},\\
  &\lbr \dz \Lambda_{1R}-\dR \Lambda_{1z} \rbr= \frac{1}{|\dl\psi_0|^2}\lbr \{\psi_0,\psi_1\}+\dl\psi_0\cdot \dl \Lambda_1 \rbr + \psi_1 \left\{\psi_0,\frac{1}{|\dl\psi_0|^2}\right\}\\
  &\qquad\qquad\qquad\qquad +\Lambda_1 \lbr \dl\psi_0 \cdot\dl \frac{1}{|\dl\psi_0|^2} + \frac{1}{|\dl\psi_0|^2}(\dz^2 +\dr^2)\psi_0\rbr.\nonumber
\end{align}
\label{eq:Lambda_del_ids}
\end{subequations}
\subsection{Group II}

Using \eqref{eq:Lambda_del_ids}, we can rewrite the Group II.1, $\dl\cdot\u_1=0$, equation as 
\begin{align}
    \del_\phi\Lambda_{1\phi}-\frac{R}{|\dl\psi_0|^2}\dl\psi_1\cdot \dl \psi_0=-\frac{R}{|\dl\psi_0|^2}\{\psi_0,\Lambda_1\}+ \cA_1 \psi_1+ \cA_2\Lambda_1\label{eq:II.1}
\end{align}
Similarly, II.2, yields
\begin{align}
    -\del_\phi\Lambda_{1\phi}+\frac{R}{(R B_0)^2}\dl\psi_1\cdot \dl \psi_0= \cA_3 \psi_1+ \cA_4\Lambda_1\label{eq:II.2}\\
    \cA_3= \frac{R}{(R B_0)^2}\lbr -\frac{(R B_0)^2}{|\dl \psi_0|^2}\frac{1}{R}\del_R \psi_0 +\frac{\dl\psi_0\cdot\dl |\dl\psi_0|^2}{2|\dl\psi_0|^2}\rbr ,\nonumber\\
    \cA_4 = \frac{1}{|\dl\psi_0|^2}\lbr \del_z \psi_0 -\frac{R}{(R B_0)^2}\left\{\psi_0,\frac{|\dl\psi_0|^2}{2}\right\} \rbr. \nonumber
\end{align}
Therefore, we find that $(u_{1\phi}=\del_\phi\Lambda_{1\phi}, \dl\psi_0\cdot \dl \psi_1)$ satisfy
\begin{align}
    \begin{pmatrix}
        \del_\phi\Lambda_{1\phi}\\
        \dl\psi_0\cdot \dl \psi_1
    \end{pmatrix}
    &=\tnsr{M_{II}}\left(
    -\frac{R}{|\dl\psi_0|^2}\{\psi_0,\Lambda_1\}
    \begin{pmatrix}
        1\\0
    \end{pmatrix}+
     \begin{pmatrix}
        \cA_1 \qquad \cA_2\\
        \cA_3 \qquad \cA_4
    \end{pmatrix}
     \begin{pmatrix}
        \psi_1\\\Lambda_1
    \end{pmatrix}
    \right)\nonumber\\
   \tnsr{M_{II}}&\equiv -\frac{|\dl\psi_0|^2 (R B_0)^2}{R F^2}
    \begin{pmatrix}
        \frac{R}{(RB_0)^2} \quad \frac{R}{|\dl\psi_0|^2}\\
        1 \qquad 1
    \end{pmatrix}.
    \label{eq:II_and_MII}
\end{align}
More explicitly,
\begin{subequations}
    \begin{align}
    &\del_\phi\Lambda_{1\phi}=\frac{R}{F^2}\{\psi_0,\Lambda_1\}+ \hat{\cA}_1 \psi_1+ \hat{\cA}_2\Lambda_1 \label{eq:u1phi_sol}\\
    &\frac{F^2}{(RB_0)^2}\dl\psi_0\cdot \dl \psi_1=\{\psi_0,\Lambda_1\}-\frac{|\dl\psi_0|^2}{R}\lbr (\cA_1+\cA_3)\psi_1 + (\cA_2+\cA_4)\Lambda_1 \rbr \label{eq:gradpsi1_sol}\\
    &\hat{\cA}_1\equiv-\frac{1}{F^2}\lbr (R B_0)^2 \cA_3+|\dl\psi_0|^2 \cA_1\rbr,\;\;
    \hat{\cA}_2\equiv-\frac{1}{F^2}\lbr (R B_0)^2 \cA_4+|\dl\psi_0|^2 \cA_2\rbr. \label{eq:A1_A2_hats}
\end{align}
\label{eq:II_sol}
\end{subequations}
Operating on \eqref{eq:u1phi_sol} by $\del_\phi$ and using the \eqref{eq:N^2_id}  we find
\begin{align}
    -N^2 \Lambda_{1\phi}=
    \frac{R}{F^2}\{\psi_0,\del_\phi\Lambda_1\}+ \hat{\cA}_1 \del_\phi\psi_1+ \hat{\cA}_2\del_\phi\Lambda_1.
    \label{eq:Lambda1phi_sol}
\end{align}

\subsection{Group III}

Equation III.1 representing $\dl\cdot\B_1=0$ can be expressed as
\begin{align}
    \lbr \frac{1}{R^2}\del_\phi +\frac{2F}{R^2}\frac{\del_z \psi_0}{|\dl\psi_0|^2}\rbr\Lambda_1-\frac{2F \del_R \psi_0}{R^2|\dl\psi_0|^2}\psi_1+\lbr \frac{2F}{R}\del_\phi+\{\psi_0,\;\; \} \rbr\lbr \frac{\Lambda_{1\phi}}{R}\rbr =0.
     \label{eq:III.1}
\end{align}
Substituting $\Lambda_{1\phi}$ from \eqref{eq:Lambda1phi_sol} we get a PDE that connects $\psi_1,\Lambda_1$ and their on-surface derivatives.


Next, we present the alternative form of Group III equations obtained from $\u_0\cdot \tnsr{S_1}\cdot \B_0$ and $\dl\psi_0\cdot \tnsr{S_1}\cdot \B_0$ respectively,
\begin{subequations}
\begin{align}
    \del_\phi\lbr \frac{1}{R}\lbr F \Lambda_{1\phi}+\Lambda_1\rbr \rbr + \B_0\cdot\dl (R \Lambda_{1\phi}) +\frac{2F}{R}\Lambda_{1R}-\frac{2}{R}(\del_z\psi_0)\Lambda_{1\phi}=0,\label{eq:III.1.alt}\\
    \dl\psi_0\cdot \dl\lbr \frac{1}{R}\lbr F \Lambda_{1\phi}+\Lambda_1\rbr \rbr - \B_0\cdot\dl \psi_1 -\frac{\dlts\psi_0}{R}\Lambda_1-\frac{F'}{R}|\dl\psi_0|^2\Lambda_{1\phi}=0. \label{eq:III.2}
\end{align}
    \label{eq:psiSB_uSB_alt_L_form}
\end{subequations}
We can rewrite \eqref{eq:III.2} as 
\begin{align}
    F \dl\psi_0\cdot \dl \lbr \frac{\Lambda_{1\phi}}{R}\rbr + \frac{1}{R}\dl\psi_0\cdot \dl \Lambda_1 -\B_0\cdot \dl \psi_1 -\frac{1}{R}\Lambda_1\lbr \dlts\psi_0 +\frac{1}{R}\del_R \psi_0\rbr=0,
    \label{eq:radial_u1phi}
\end{align}
i.e., an equation determining the radial gradient of $\Lambda_{1\phi}$ in terms of $\psi_1$ and $\Lambda_1$ and their derivatives. 

By eliminating the $F \Lambda_{1\phi}+\Lambda_1$ terms from \eqref{eq:psiSB_uSB_alt_L_form}, we obtain a consistency condition
\begin{align}
    &\B_0\cdot \dl \lbr \del_\phi\psi_1 +\dl\psi_0\cdot \dl (R\Lambda_{1\phi}) \rbr +\frac{1}{R}\del_\phi\lbr \dlts\psi_0  \Lambda_1+F' |\dl\psi_0|^2\Lambda_{1\phi}\rbr \nonumber\\
    &+\left[\dl\psi_0\cdot \dl, \B_0\cdot \dl\right](R\Lambda_{1\phi}) +\dl\psi_0\cdot \dl \lbr \frac{2}{R}\lbr F \Lambda_{1R}-\del_z\psi_0 \Lambda_{1\phi}\rbr \rbr=0,
    \label{eq:towards_shear_condn}
\end{align}
where $[\;,\;]$ denotes the standard commutator of two operators. The flux-surface average of \eqref{eq:towards_shear_condn} determines the magnetic shear self-consistent with quasisymmetry. 


\subsection{Group IV}
The final step is to simplify the GGSE \eqref{eq:GGS_cyl_L} and express all quantities in terms of $\psi_1$ and $\Lambda_1$ and their derivatives. It is simpler to look at the GGSE for $\Psi_{1}\equiv (1/N)\del_\phi \psi_1$,
given by
\begin{align}
    &\dlts \Psi_1 +\Psi_1 \lbr {F'}^2 +F F'' + R^2 p''\rbr +\frac{1}{N}\del_\phi v_{1B} \nonumber\\
    &\quad-\frac{2}{R^2}N\lbr R\Lambda_{1\phi}\lbr R^2 p' +\frac{\del_R \psi_0}{R}\rbr +F \Lambda_{1z} \rbr=0, \nonumber\\
    &\frac{1}{N}\del_\phi v_{1B}  =  -N\left[\frac{F}{R}\lbr \del_z \Lambda_{1R}-\del_R \Lambda_{1z}\rbr -\frac{1}{R^2}\dl\psi_0\cdot \dl (R\Lambda_{1\phi})\right].
    \label{eq:Dlts_Psi_1_form_of_GGSE}
\end{align}

To bring the GGSE to a form which only involves $\psi_1, \Lambda_1$ and their derivatives, we now substitute the expressions and identities \eqref{eq:Lambda_del_ids} for $\Lambda_i$'s into $v_{1B}$ to get
\begin{align}
    -\frac{1}{N^2}\del_\phi v_{1B}=& \frac{F}{R}\left\{\psi_0,\frac{\psi_1}{|\dl\psi_0|^2}\right\}
    +\frac{F}{R}\frac{\dl\psi_0\cdot\dl\Lambda_1}{|\dl\psi_0|^2}-\dl\psi_0\cdot \dl\lbr \frac{\Lambda_{1\phi}}{R} \rbr\\
    &+\frac{F}{R}\Lambda_1\lbr \del_z\lbr\frac{\del_z\psi_0}{|\dl\psi_0|^2}\rbr +\del_R\lbr\frac{\del_R\psi_0}{|\dl\psi_0|^2}\rbr\rbr-\frac{2}{R^2}\del_R \psi_0 \Lambda_{1\phi}. \nonumber
\end{align}
Upon using the equation for $\dl\psi_0\cdot\dl (\Lambda_{1\phi}/R)$ given in \eqref{eq:radial_u1phi} and simplifying, we find
\begin{align}
    -\frac{1}{N^2}\del_\phi v_{1B}= \frac{F}{R}\frac{1}{|\dl\psi_0|^2}\lbr \left\{\psi_0,\psi_1\right\} -\psi_1 \frac{1}{|\dl\psi_0|^2}\{\psi_0,|\dl\psi_0|^2\} \rbr-\frac{1}{F}\B_0\cdot\dl\psi_1\nonumber\\
    +\frac{F}{R |\dl\psi_0|^2}\Lambda_1\lbr \lbr\dlts\psi_0+\frac{\del_R\psi_0}{R}\rbr \lbr 1 -\lbr\frac{|\dl\psi_0|}{F}\rbr^2\rbr-2\cA_0\rbr \nonumber\\
    +\frac{F}{R}\frac{\dl\psi_0\cdot\dl\Lambda_1}{|\dl\psi_0|^2} \lbr 1 +\lbr\frac{|\dl\psi_0|}{F}\rbr^2\rbr-\frac{2}{R^2}\del_R \psi_0 \Lambda_{1\phi}. 
\end{align}
Assembling all the terms together in \eqref{eq:Dlts_Psi_1_form_of_GGSE}, we finally obtain
\begin{align}
    \lbr \dlts +\cC_\Psi \rbr \Psi_1 &-\frac{N F}{R|\dl\psi_0|^2}\{\psi_0,\psi_1\}+\frac{N F}{F^2}\B_0\cdot\dl\psi_1+N\lbr \cC_\psi \psi_1 +\cC_\Lambda \Lambda_1 +\cC_\phi \Lambda_{1\phi}\rbr\nonumber\\
    &=\frac{N}{RF}\frac{(R B_0)^2}{|\dl\psi_0|^2} \dl\psi_0 \cdot \dl \Lambda_1, 
    \label{eq:GGSE_comp}
\end{align}
where,
\begin{align}
   &\cC_\Psi={F'}^2 +F F'' + R^2 p'',\quad \cC_\phi = -\frac{2}{R}R^2 p',\quad \cA_0 = \frac{\dl\psi_0\cdot \dl |\dl\psi_0|^2}{ 2|\dl\psi_0|^2},\nonumber\\
    &\cC_\psi=\frac{ F}{R|\dl\psi_0|^2}\lbr\frac{1}{|\dl\psi_0|^2}\left\{\psi_0, |\dl\psi_0|^2\right\} +\frac{2}{R}\del_z \psi_0\rbr, \nonumber\\
    &\cC_{\Lambda}=-\frac{ F}{R|\dl\psi_0|^2}\left[\lbr 1 -\lbr\frac{|\dl\psi_0|}{F}\rbr^2\rbr \lbr \dlts\psi_0+\frac{\del_R\psi_0}{R}\rbr -2\cA_0 -\frac{2}{R}\del_R \psi_0\right]. \nonumber
\end{align}

The expressions for the various coefficients that appear in the NASE equations are collected below
\begin{align}
    &\cA_0 = \frac{\dl\psi_0\cdot \dl |\dl\psi_0|^2}{ 2|\dl\psi_0|^2}, \quad
    \cA_1 =\frac{R}{|\dl \psi_0|^2}\lbr \nabla^2 \psi_0 -2\cA_0\rbr, \label{eq:AS_coeffs}\\
    &\cA_2 = -\left\{\psi_0,\frac{R}{|\dl \psi_0|^2}\right\},\quad 
    \cA_3= \frac{R}{(R B_0)^2} \cA_0 -\frac{\del_R \psi_0}{|\dl \psi_0|^2}  ,\nonumber\\
    &\cA_4 = \frac{1}{|\dl\psi_0|^2}\lbr \del_z \psi_0 -\frac{R}{(R B_0)^2}\left\{\psi_0,\frac{|\dl\psi_0|^2}{2}\right\} \rbr, \nonumber\\
    &\hat{\cA}_1=-\frac{1}{F^2}\lbr (R B_0)^2 \cA_3+|\dl\psi_0|^2 \cA_1\rbr,\;\;
    \hat{\cA}_2=-\frac{1}{F^2}\lbr (R B_0)^2 \cA_4+|\dl\psi_0|^2 \cA_2\rbr , \nonumber\\
    &\cC_\Psi={F'}^2 +F F'' + R^2 p'',\quad \cC_\phi = -\frac{2}{R}R^2 p', \nonumber\\
    &\cC_\psi=\frac{ F}{R|\dl\psi_0|^2}\lbr\frac{1}{|\dl\psi_0|^2}\left\{\psi_0, |\dl\psi_0|^2\right\} +\frac{2}{R}\del_z \psi_0\rbr, \nonumber\\
    &\cC_\Lambda=-\frac{ F}{R|\dl\psi_0|^2}\left[\lbr 1 -\lbr\frac{|\dl\psi_0|}{F}\rbr^2\rbr \lbr \dlts\psi_0+\frac{\del_R\psi_0}{R}\rbr -2\cA_0 -\frac{2}{R}\del_R \psi_0\right]. \nonumber
\end{align}

In summary, we find that the first-order NASE system \eqref{eq:O(ep))_cyl_L} can be rearranged into Groups I-IV. Group I, consisting of \eqref{eq:uDpsi_cyl_L}, and the definition of $\Lambda_1$ \eqref{eq:Lambda_def}, determines $\Lambda_{1R}$ and $\Lambda_{1z}$. Group II made up of \eqref{eq:div_u_cyl_L} and \eqref{eq:uDB_cyl_L}, determines $\Lambda_{1\phi}, \dl\psi_0\cdot\dl \psi_1$. Group III formed from \eqref{eq:div_B_cyl_L} and \eqref{eq:JwB_cyl_L}, yields the on-surface equation \eqref{eq:III.1} for $(\psi_1,\Lambda_1)$. It also yields an equation for the radial gradient of $\Lambda_{1\phi}$ \eqref{eq:radial_u1phi}. Since $\Lambda_{1\phi}$ is already related to $(\psi_1,\Lambda_1)$ in Group III, the additional radial derivative equation leads to a consistency condition \eqref{eq:towards_shear_condn}, which involves the magnetic shear. Finally, Group IV shows that the first-order GGSE \eqref{eq:GGS_cyl_L} furnishes a radial derivative for $\Lambda_1$.

\section{Gaussian curvature of an axisymmetric system}
\label{app:K}
Although the expression for Gaussian curvature, $\cK$, in a surface of revolution is well-known, an expression that relates $\cK$ to the Grad-Shafranov equation is not readily available. To this end, we derive the expression for $\cK$ for the axisymmetric background in this Appendix.

Let $\psi_0$ be an axisymmetric toroidal surface parameterized by poloidal angle $\theta$, and toroidal angle $\phi$. The position vector $\bm{r}\equiv(x,y,z)$ can be expressed in terms of the ``inverse coordinates" $(\psi_0,\theta,\phi)$ with $\{\psi_0,\theta\}=1$ using
\begin{align}
    x(\psi_0,\theta,\phi)= R(\psi_0,\theta)\cos\phi, \;\; y(\psi_0,\theta,\phi)= R(\psi_0,\theta)\sin\phi, \;\; z(\psi_0,\theta,\phi)= Z(\psi_0,\theta).
    \label{eq:xyz_AS}
\end{align}
The elements of the first fundamental form $I=\ \cE d\theta^2  +2 \cF d\theta d\phi +\cG d\phi^2 $ are
\begin{align}
\cE= |\del_\theta\bm{r}|^2=(\del_\theta R^2)+(\del_\theta Z^2), \quad  \cF=\del_\theta\bm{r}\cdot \del_\phi\bm{r}=0,\quad \cG= |\del_\phi\bm{r}|^2=R^2
    \label{eq:first_fund_form}
\end{align}
The normal vector $\hat{\bm{N}}=(\del_\theta \bm{r}\times \del_\phi \bm{r})/|\del_\theta \bm{r}\times \del_\phi \bm{r}|$ is of the form
\begin{align}
   \hat{\bm{N}}=\frac{1}{\sqrt{(\del_\theta R)^2+(\del_\theta Z)^2}} \lbr -\del_\theta Z \lbr \cos\phi\; \hat{\bm{x}}+\sin\phi\; \hat{\bm{y}}\rbr +\del_\theta R\; \hat{\bm{z}} \rbr
\end{align}
The elements of the second fundamental form $II= \mathfrak{e}d\theta^2   +2  \mathfrak{f} d\theta d\phi + \mathfrak{g} d\phi^2$ are
\begin{align}
    \mathfrak{e}&\equiv\hat{\bm{N}}\cdot\del_{\theta}^2\bm{r}=
    \frac{-\del_\theta Z\; \del_\theta^2 R + \del_\theta R\; \del_\theta^2 Z}{\sqrt{(\del_\theta R)^2+(\del_\theta Z)^2}}, \quad \mathfrak{f}\equiv \hat{\bm{N}}\cdot\del_{\theta}\del_{\phi}\bm{r}=0, \nonumber\\
     \mathfrak{g}&\equiv \hat{\bm{N}}\cdot\del_{\phi}^2\bm{r}=\frac{R \del_\theta Z}{\sqrt{(\del_\theta R)^2+(\del_\theta Z)^2}}.
\end{align}
The principal curvatures $\kappa_1= \mathfrak{e}/\cE,\kappa_2=\mathfrak{g}/\cG$ are 
\begin{align}
    \kappa_1 = \frac{-\del_\theta Z\; \del_\theta^2 R + \del_\theta R\; \del_\theta^2 Z}{\lbr(\del_\theta R)^2+(\del_\theta Z)^2\rbr^{3/2}}, \quad \kappa_2 = \frac{\del_\theta Z}{R{\sqrt{(\del_\theta R)^2+(\del_\theta Z)^2}}}.
\end{align}
The mean and Gaussian curvatures $\cH=(\kappa_1+\kappa_2)/2,\cK=\kappa_1\kappa_2$ are
\begin{align}
\cH &= \frac{-(\del_\theta Z\; \del^2_\theta R-\del_\theta R\; \del^2_\theta Z) +\frac{\del_\theta Z}{R}\lbr (\del_\theta R)^2+(\del_\theta Z)^2\rbr}{2{\lbr(\del_\theta R)^2+(\del_\theta Z)^2\rbr^{3/2}}},\nonumber\\
    \cK&=\frac{-\del_\theta Z(\del_\theta Z\; \del_\theta^2 R - \del_\theta R \;\del_\theta^2 Z)}{\lbr(\del_\theta R)^2+(\del_\theta Z)^2\rbr^{2}}.
    \label{eq:H_and_K}
\end{align}
To connect $\cK$ to the Grad-Shafranov equation, we need $\cK$ in terms of $\psi_0$ and its $(R,z)$ gradients, rather than in terms of the inverse coordinates as given in \eqref{eq:H_and_K}. So we change coordinates back to $(R,\phi,z)$. Note that the normal vector is simply $\hat{\bm{N}}=\dl\psi_0/|\dl\psi_0|$. It follows that
\begin{align}
    &\del_{\psi_0} R= -\del_z \theta ,\;\; \del_{\psi_0} Z= \del_R \theta ,\;\;\del_\theta R = \del_z \psi_0,\;\; \del_\theta Z = -\del_R \psi_0\\
    &\del^2_\theta R = (\del_z\psi_0)( \del_{z}\del_R\psi_0)-(\del_R\psi_0)( \del^2_{z}\psi_0) , \quad \del^2_\theta Z = (\del_R\psi_0)( \del_{z}\del_R\psi_0)-(\del_z\psi_0)( \del^2_{R}\psi_0)\nonumber
\end{align}
we find that $\cK$ can be written as
\begin{align}
    \cK = \lbr \frac{\del_R\psi_0}{R}\rbr\frac{\lbr (\del_z\psi_0)^2\del^2_R\psi_0 +(\del_R\psi_0)^2\del^2_z\psi_0 -2 (\del_R\psi_0)(\del_z \psi_0)\del_R\del_z\psi_0\rbr}{\lbr(\del_R\psi_0)^2 +(\del_z\psi_0)^2\rbr^2}.
    \label{eq:cK_exp}
\end{align}
Now, from the expression
\begin{align}
    \cA_0 =\frac{\lbr (\del_R\psi_0)^2\del^2_R\psi_0 +(\del_z\psi_0)^2\del^2_z\psi_0 +2 (\del_R\psi_0)(\del_z \psi_0)\del_R\del_z\psi_0\rbr}{\lbr(\del_R\psi_0)^2 +(\del_z\psi_0)^2\rbr},
    \label{eq:cA_0_exp}
\end{align}
we find that
\begin{align}
    \cK=\lbr \frac{\del_R\psi_0/R}{|\dl\psi_0|^2}\rbr \lbr (\del_R^2+\del_z^2)\psi_0 -\cA_0 \rbr\;\;=\frac{\delta_R}{F^2} \lbr 2\delta_J +\delta_R \vep_p^2-\cA_0\rbr.
\end{align}

\bibliographystyle{jpp}
\bibliography{plasmalit}

@article{aronsson1968partial,
  title={On the partial differential equation $u_x^2 u_{xx}+ 2 u_x u_y u_{xy}+ u_y^2 u_{yy}= 0$},
  author={Aronsson, G.},
  journal={Arkiv f{\"o}r matematik},
  volume={7},
  pages={395--425},
  year={1968},
  publisher={Springer}
}

@article{aronsson_crandell2004tour,
  title={A tour of the theory of absolutely minimizing functions},
  author={Aronsson, G. and Crandall, M. and Juutinen, P.},
  journal={Bulletin of the American mathematical society},
  volume={41},
  number={4},
  pages={439--505},
  year={2004}
}

@article{bader2019,
  title     = {Stellarator equilibria with reactor relevant energetic particle losses},
  volume    = {85},
  doi       = {10.1017/S0022377819000680},
  number    = {5},
  journal   = {Journal of Plasma Physics},
  publisher = {Cambridge University Press},
  author    = {Bader, A. and Drevlak, M. and Anderson, D. T. and Faber, B. J. and Hegna, C. C. and Likin, K. M. and Schmitt, J. C. and Talmadge, J. N.},
  year      = 2019,
  pages     = {905850508}
}

@article{berry1976waves_and_Thom,
  title={Waves and {Thom's} theorem},
  author={Berry, M. V.},
  journal={Advances in Physics},
  volume={25},
  number={1},
  pages={1--26},
  year={1976},
  publisher={Taylor \& Francis}
}

@article{boozer_2015, 
  title={Stellarator design}, 
  volume={81}, 
  number={6}, 
  journal={Journal of Plasma Physics}, 
  publisher={Cambridge University Press}, 
  author={Boozer, A. H.}, 
  year={2015}, 
  pages={515810606}
}

@article{burby2020,
  author  = {Burby, J. W. and Kallinikos, N. and MacKay, R. S.},
  title   = {Some mathematics for quasi-symmetry},
  journal = {Journal of Mathematical Physics},
  volume  = {61},
  number  = {9},
  pages   = {093503},
  year    = {2020},
  doi     = {10.1063/1.5142487}
}

@article{burby2024characterization,
  title={Characterization of admissible quasisymmetries},
  author={Burby, J. W. and Kallinikos, N. and MacKay, R. S.},
  journal={Journal of Mathematical Physics},
  volume={65},
  number={6},
  pages={061502},
  year={2024},
  publisher={AIP Publishing}
}

@article{constantin2021,
  title     = {On quasisymmetric plasma equilibria sustained by small force},
  volume    = {87},
  doi       = {10.1017/S0022377820001610},
  number    = {1},
  journal   = {Journal of Plasma Physics},
  publisher = {Cambridge University Press},
  author    = {Constantin, P. and Drivas, T. D. and Ginsberg, D.},
  year      = {2021},
  pages     = {905870111}
}

@book{eisenhart1923transformations,
  title={Transformations of surfaces},
  author={Eisenhart, L. P.},
  year={1923},
  publisher={Princeton University Press}
}

@book{forsyth_v4_pt6,
  title={Theory of Differential Equations, Part IV, Partial Differential Equations},
  author={Forsyth, A. R.},
  volume={{VI}},
  year={1906},
  publisher={Cambridge University Press}
}

@book{freidberg2014idealMHD,
  title={Ideal {MHD}},
  author={Freidberg, J. P.},
  year={2014},
  publisher={Cambridge University Press},
  pages={302}
}

@article{dewar1983ballooning,
  title={Ballooning mode spectrum in general toroidal systems},
  author={Dewar, R. L. and Glasser, A. H.},
  journal={The Physics of fluids},
  volume={26},
  number={10},
  pages={3038--3052},
  year={1983},
  publisher={American Institute of Physics}
}

@article{garcia2024resilient,
  title={Resilient stellarator divertor characteristics in the helically symmetric eXperiment},
  author={Garcia, K. A. and Bader, A. and Boeyaert, D. and Boozer, A. H. and Frerichs, H. and Gerard, M. J. and Punjabi, A. and Schmitz, O.},
  journal={Plasma Physics and Controlled Fusion},
  volume={67},
  number={3},
  pages={035011},
  year={2025},
  publisher={IOP Publishing}
}

@article{garcia2023explorationCTH,
  title={Exploration of non-resonant divertor features on the Compact Toroidal Hybrid},
  author={Garcia, K. A. and Bader, A. and Frerichs, H. and Hartwell, G. J. and Schmitt, J. C. and Allen, N. and Schmitz, O.},
  journal={Nuclear Fusion},
  volume={63},
  number={12},
  pages={126043},
  year={2023},
  publisher={IOP Publishing}
}

@article{helander2014,
  title={Theory of plasma confinement in non-axisymmetric magnetic fields},
  author={Helander, P.},
  journal={Reports on Progress in Physics},
  volume={77},
  number={8},
  pages={087001},
  year={2014},
  publisher={IOP Publishing}
}

@article{henneberg2024compact,
  title={Compact stellarator-tokamak hybrid},
  author={Henneberg, S. A. and Plunk, G. G.},
  journal={Physical Review Research},
  volume={6},
  number={2},
  pages={L022052},
  year={2024},
  publisher={APS}
}

@article{Schuett_henneberg2024compactQA,
  title={Exploring novel compact quasi-axisymmetric stellarators},
  author={Schuett, T. M. and Henneberg, S. A.},
  journal={Physical Review Research},
  volume={6},
  pages={L042052},
  year={2024},
  publisher={APS}
}

@article{landremanBullerDrevlak2022optimization,
  title={Optimization of quasi-symmetric stellarators with self-consistent bootstrap current and energetic particle confinement},
  author={Landreman, M. and Buller, S. and Drevlak, M.},
  journal={Physics of Plasmas},
  volume={29},
  number={8},
  pages={082501},
  year={2022},
  publisher={AIP Publishing}
}

@article{nikulsin_sengupta2024GS,
  title={An asymptotic {Grad-Shafranov} equation for quasisymmetric stellarators},
  author={Nikulsin, N. and Sengupta, W. and Jorge, R. and Bhattacharjee, A.},
  journal={Journal of Plasma Physics},
  volume={90},
  number={6},
  pages={905900608},
  year={2024},
  doi={10.1017/S0022377824000916},
  publisher={Cambridge University Press}
}

@article{plunk2018,
  title     = {Quasi-axisymmetric magnetic fields: weakly non-axisymmetric case in a vacuum},
  volume    = {84},
  doi       = {10.1017/S0022377818000259},
  number    = {2},
  journal   = {Journal of Plasma Physics},
  publisher = {Cambridge University Press},
  author    = {Plunk, G. G. and Helander, P.},
  year      = {2018},
  pages     = {905840205}
}

@article{plunk2020_near_axisymmetry_MHD,
  title={Perturbing an axisymmetric magnetic equilibrium to obtain a quasi-axisymmetric stellarator},
  author={Plunk, G. G.},
  journal={Journal of Plasma Physics},
  volume={86},
  number={4},
  pages={905860409},
  year={2020},
  publisher={Cambridge University Press}
}

@article{rodrigGBC,
  author  = {Rodr{\'\i}guez, E. and Sengupta, W. and Bhattacharjee, A.},
  title   = {Generalized {Boozer} coordinates: A natural coordinate system for quasisymmetry},
  journal = {Physics of Plasmas},
  volume  = {28},
  number  = {9},
  pages   = {092510},
  year    = {2021},
  doi     = {10.1063/5.0060115}
}

@article{rodriguez2020a,
  author  = {Rodr{\'\i}guez, E. and Helander, P. and Bhattacharjee, A.},
  title   = {Necessary and sufficient conditions for quasisymmetry},
  journal = {Physics of Plasmas},
  volume  = {27},
  number  = {6},
  pages   = {062501},
  year    = {2020},
  doi     = {10.1063/5.0008551}
}

@article{rodriguez2021weak,
  author  = {Rodr{\'\i}guez, E. and Sengupta, W. and Bhattacharjee, A.},
  title   = {Weakly quasisymmetric near-axis solutions to all orders},
  journal = {Physics of Plasmas},
  volume  = {29},
  number  = {1},
  pages   = {012507},
  year    = {2022},
  doi     = {10.1063/5.0076583}
}

@article{sengupta2021NSE,
  title={Vacuum magnetic fields with exact quasisymmetry near a flux surface. Part 1. Solutions near an axisymmetric surface},
  author={Sengupta, W. and Paul, E. J. and Weitzner, H. and Bhattacharjee, A.},
  journal={Journal of Plasma Physics},
  volume={87},
  number={2},
  pages={905870205},
  year={2021},
  publisher={Cambridge University Press}
}

@article{sengupta2024NAE_finite_beta,
  title={Stellarator equilibrium axis-expansion to all orders in distance from the axis for arbitrary plasma beta},
  author={Sengupta, W. and Rodriguez, E. and Jorge, R. and Landreman, M. and Bhattacharjee, A.},
  journal={Journal of Plasma Physics},
  volume={90},
  number={4},
  pages={905900412},
  year={2024},
  publisher={Cambridge University Press}
}

@article{sengupta2024QSHBS,
  title={Asymptotic quasisymmetric high-beta three-dimensional magnetohydrodynamic equilibria near axisymmetry},
  author={Sengupta, W. and Nikulsin, N. and Gaur, R. and Bhattacharjee, A.},
  journal={Journal of Plasma Physics},
  volume={90},
  number={2},
  pages={905900209},
  year={2024},
  publisher={Cambridge University Press}
}

@article{strauss1997reduced_near_vacuum,
  title={Reduced {MHD} in nearly potential magnetic fields},
  author={Strauss, H. R.},
  journal={Journal of Plasma Physics},
  volume={57},
  number={1},
  pages={83--87},
  year={1997},
  publisher={Cambridge University Press}
}

@article{nuhrenberg2006critical,
  title={Critical issues and comparison of optimized stellarators},
  author={N{\"u}hrenberg, J.},
  journal={Fusion science and technology},
  volume={50},
  number={2},
  pages={146--157},
  year={2006},
  publisher={Taylor \& Francis}
}

@article{bader2017_resilient_div_hsx,
  title={{HSX} as an example of a resilient non-resonant divertor},
  author={Bader, A. and Boozer, A. H. and Hegna, C. C. and Lazerson, S. A. and Schmitt, J. C.},
  journal={Physics of Plasmas},
  volume={24},
  number={3},
  pages={032506},
  year={2017},
  publisher={AIP Publishing}
}

@article{strumberger1996sol,
  title={{SOL} studies for {W7-X} based on the island divertor concept},
  author={Strumberger, E.},
  journal={Nuclear fusion},
  volume={36},
  number={7},
  pages={891},
  year={1996},
  publisher={IOP Publishing}
}

@article{buller2024family,
  title={A family of quasi-axisymmetric stellarators with varied rotational transform},
  author={Buller, S. and Landreman, M. and Kappel, J. and Gaur, R.},
  journal={Journal of Plasma Physics},
  volume={90},
  number={1},
  pages={905900115},
  year={2024},
  publisher={Cambridge University Press}
}

@article{pataki_cerfon_2013fastGS,
  title={A fast, high-order solver for the Grad--Shafranov equation},
  author={Pataki, A. and Cerfon, A. J. and Freidberg, J. P. and Greengard, L. and O’Neil, M.},
  journal={Journal of Computational Physics},
  volume={243},
  pages={28--45},
  year={2013},
  publisher={Elsevier}
}

@article{landreman2021simsopt,
  title={SIMSOPT: a flexible framework for stellarator optimization},
  author={Landreman, M. and Medasani, B. and Wechsung, F. and Giuliani, A. and Jorge, R. and Zhu, C.},
  journal={Journal of Open Source Software},
  volume={6},
  number={65},
  pages={3525},
  year={2021}
}

@book{magnus2013hill_eq,
  title={Hill's equation},
  author={Magnus, W. and Winkler, S.},
  year={2013},
  publisher={Courier Corporation}
}

@book{coddington1956theoryODE,
  title={Theory of ordinary differential equations},
  author={Coddington, E. A. and Levinson, N.},
  year={1955},
  publisher={McGraw-Hill}
}

@article{brown2025Palumbo,
  title={A class of high-beta, large-aspect-ratio quasiaxisymmetric Palumbo-like configurations},
  author={Brown, A. and Sengupta, W. and Nikulsin, N. and Bhattacharjee, A.},
  journal={arXiv preprint arXiv:2506.17528},
  year={2025}
}

@article{virtanen2020scipy,
  author = {Virtanen, P. and Gommers, R. and Oliphant, T. E. and Haberland, M. and Reddy, T. and Cournapeau, D. and Burovski, E. and Peterson, P. and Weckesser, W. and Bright, J. and van der Walt, S. J. and Brett, M. and Wilson, J. and Millman, K. J. and Mayorov, N. and Nelson, A. R. J. and Jones, E. and Kern, R. and Larson, E. and Carey, C. J. and Polat, I. and Feng, Y. and Moore, E. W. and VanderPlas, J. and Laxalde, D. and Perktold, J. and Cimrman, R. and Henriksen, I. and Quintero, E. A. and Harris, C. R. and Archibald, A. M. and Ribeiro, A. H. and Pedregosa, F. and van Mulbregt, P. and {SciPy 1.0 Contributors}},
  title = {{{SciPy} 1.0: Fundamental Algorithms for Scientific Computing in Python}},
  journal = {Nature Methods},
  volume = {17},
  pages = {261--272},
  year = {2020},
  doi = {10.1038/s41592-019-0686-2}
}

@book{bettess1992infinite,
  author = {Bettess, P.},
  publisher = {Penshaw Press},
  title = {Infinite elements},
  edition = {1},
  year = {1992},
  url = {https://worldcat.org/title/1020709342}
}

@article{PaperI,
  title={Optical analogy for stellarators: Ridges as caustics and coils as singularities},
  author={Sengupta, W. and Buller, S. and Jorge, R. and Kappel, J. and Brown, A. and Nies, R. and Gil, P. F. and Nikulsin, N. and Helander, P. and Bhattacharjee, A.},
  journal = {preprint arXiv:2605.21814},
  year={2026}
}

@article{nikulsin2025a,
  author  = {Nikulsin, N. and Sengupta, W. and Buller, S. and Bhattacharjee, A.},
  title   = {A {Grad-Shafranov} model for compact quasisymmetric stellarators},
  journal = {Physics of Plasmas},
  volume  = {32},
  number  = {4},
  pages   = {042501},
  year    = {2025},
  doi     = {10.1063/5.0258308}
}

\end{document}